\documentclass[aps,pra,twocolumn,showpacs]{revtex4}
\usepackage{amsmath}
\usepackage{amssymb}
\usepackage{graphicx}
\usepackage{longtable}

\def\refeqn#1{Eq.\ (\ref{Equation::#1})}
\def\refeqns#1#2{Eqs.\ (\ref{Equation::#1}) and (\ref{Equation::#2})}

\def\refsec#1{Section \ref{Section::#1}}

\def\refapx#1{Appendix \ref{Appendix::#1}}

\def\reffig#1{Figure \ref{Figure::#1}}

\def\reftab#1{Table \ref{Table::#1}}
\def\reftabs#1#2{Tables \ref{Table::#1} and \ref{Table::#2}}

\begin{document}
    \title{Robust quantum parameter estimation: coherent magnetometry with feedback}
    \author{John K. Stockton}
    \email{jks@caltech.edu}
    \author{JM Geremia}
    \author{Andrew C. Doherty}
    \author{Hideo Mabuchi}
    \affiliation{Norman Bridge Laboratory of Physics, M.C.
    12-33, California Institute of Technology, Pasadena CA 91125}


\date{\today}

\begin{abstract}

We describe the formalism for optimally estimating and controlling
both the state of a spin ensemble and a scalar magnetic field with
information obtained from a continuous quantum limited measurement
of the spin precession due to the field.  The full quantum
parameter estimation model is reduced to a simplified equivalent
representation to which classical estimation and control theory is
applied.  We consider both the tracking of static and fluctuating
fields in the transient and steady state regimes.  By using
feedback control, the field estimation can be made robust to
uncertainty about the total spin number.

\end{abstract}

\pacs{07.55.Ge,03.65.Ta,42.50.Lc,02.30.Yy}

\maketitle

\section{Introduction}\label{Section::Introduction}

As experimental methods for manipulating physical systems near
their fundamental quantum limits improve \cite{Armen2002,
Geremia2003b, Orozco2002, Raithel2002, Rempe2002}, the need for
quantum state and parameter estimation methods becomes critical.
Integrating a modern perspective on quantum measurement theory
with the extensive methodologies of classical estimation and
control theory provides new insight into how the limits imposed by
quantum mechanics affect our ability to measure and control
physical systems \cite{Verstraete2001, Gambetta2001, Mabuchi1996,
Belavkin1999}.

In this paper, we illustrate the processes of state estimation and
control for a continuously-observed, coherent spin ensemble (such
as an optically pumped cloud of atoms) interacting with an
external magnetic field.  In the situation where the magnetic
field is either zero or well-characterized, continuous measurement
(e.g., via the dispersive phase shift or Faraday rotation of a far
off-resonant probe beam) can produce a spin-squeezed
\cite{Kitagawa1993} state conditioned on the measurement record
\cite{Kuzmich2000}.  Spin-squeezing indicates internal
entanglement between the different particles in the ensemble
\cite{Stockton2003} and promises to improve precision measurements
\cite{Wineland1994}.  When, however, the ambient magnetic
environment is either unknown or changing in time, the external
field can be estimated by observing Larmor precession in the
measurement signal \cite{Geremia2003b, Jessen2003, Romalis2003,
Budker2002}, see \reffig{Schematic}. Recently, we have shown that
uncertainty in both the magnetic field and the spin ensemble can
be simultaneously reduced through continuous measurement and
adequate quantum filtering \cite{Geremia2003}.

Here, we expand on our recent results \cite{Geremia2003} involving
Heisenberg-limited magnetometry by demonstrating the advantages of
including feedback control in the estimation process.  Feedback is
a ubiquitous concept in classical applications because it enables
precision performance despite the presence of potentially large
system uncertainty.  Quantum optical experiments are evolving to
the point where feedback can been used, for example, to stabilize
atomic motion within optical lattices \cite{Raithel2002} and high
finesse cavities \cite{Rempe2002}. Recently, we demonstrated the
use of feedback on a polarized ensemble of laser-cooled Cesium
atoms to robustly estimate an applied magnetic field
\cite{Geremia2003b}. In this work, we investigate the theoretical
limits of such an approach and demonstrate that an external
magnetic field can be measured with high precision despite
substantial ignorance of the size of the spin ensemble.

The paper is organized as follows. In
\refsec{QuantumParameterEstimation}, we provide a general
introduction to quantum parameter estimation followed by a
specialization to the case of a continuously measured spin
ensemble in a magnetic field.  By capitalizing on the Gaussian
properties of both coherent and spin-squeezed states, we formulate
the parameter estimation problem in such a way that techniques
from classical estimation theory apply to the quantum system.
\refsec{OptimalEstimationAndControl} presents basic filtering and
control theory in a pedagogical manner with the simplified spin
model as an example. This theory is applied in
\refsec{OptimalPerformance}, where we simultaneously derive
mutually dependent magnetometry and spin-squeezing limits in the
ideal case where the observer is certain of the spin number. We
consider the optimal measurement of both constant and fluctuating
fields in the transient and steady state regimes. Finally, we show
in \refsec{RobustPerformance} that the estimation can be made
robust to uncertainty about the total spin number by using
precision feedback control.

\begin{figure*}
\includegraphics[width=6.5in]{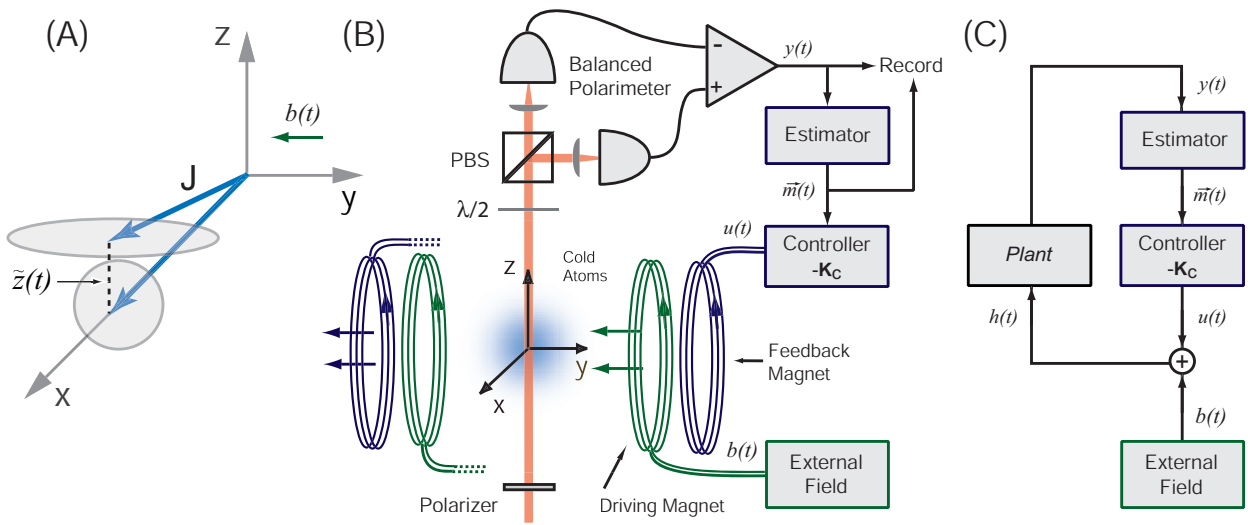} \caption{
(A) A spin ensemble is initially prepared in a coherent state
polarized along x, with symmetric variance in the y and z
directions.  Subsequently, a field along y causes the spin to
rotate as the z-component is continuously measured.  (B)
Experimental schematic for the measurement process.  A far off
resonant probe beam traverses the sample and measures the
z-component of spin via Faraday rotation. The measurement strength
could be improved by surrounding the ensemble with a cavity. (C)
Experimental apparatus subsumed by the \textit{Plant} block, which
serves to map the total field to the photocurrent,
$h(t)\rightarrow y(t)$.}\label{Figure::Schematic}
\end{figure*}

\section{Quantum parameter estimation}\label{Section::QuantumParameterEstimation}

First, we present a generic description of quantum parameter
estimation \cite{Verstraete2001, Gambetta2001, Mabuchi1996,
Belavkin1999}.  This involves describing the quantum system with a
density matrix and our knowledge of the unknown parameter with a
classical probability distribution.   The objective of parameter
estimation is then to utilize information gained about the system
through measurement to conditionally update both the density
matrix and the parameter distribution.   After framing the general
case, our particular example of a continuously measured spin
ensemble is introduced.

\subsection{General problem}

The following outline of the parameter estimation process could be
generalized to treat a wide class of problems (discrete
measurement, multiple parameters), but for simplicity, we will
consider a continuously measured quantum system with scalar
Hamiltonian parameter $\theta$ and measurement record $y(t)$.

Suppose first that the observer has full knowledge of the
parameter $\theta$. The proper description of the system would
then be a density matrix $\rho_\theta(t)$ conditioned on the
measurement record $y(t)$. The first problem is to find a rule to
update this density matrix with the knowledge obtained from the
measurement. As in the problem of this paper, this mapping may
take the form of a stochastic master equation (SME). The SME is by
definition a filter that maps the measurement record to an optimal
estimate of the system state.

Now if we allow for uncertainty in $\theta$, then a particularly
intuitive choice for our new description of the system is
\begin{eqnarray}
\rho(t)&\equiv&\int_{\theta}\rho_\theta(t)p(\theta,t)\,d\theta\label{Equation::FullEstimate}
\end{eqnarray}
where $p(\theta,t)$ is a probability distribution representing our
knowledge of the system parameter.  In addition to the rule for
updating each $\rho_\theta(t)$, we also need to find a rule for
updating $p(\theta,t)$ according to the measurement record. By
requiring internal consistency, it is possible to find a Bayes
rule for updating $p(\theta,t)$ \cite{Verstraete2001}. These two
update rules in principle solve the estimation problem completely.

Because evolving $\rho(t)$ involves performing calculations with
the full Hilbert space in question, which is often computationally
expensive, it is desirable to find a reduced description of the
system. Fortunately, it is often possible to find a closed set of
dynamical equations for a small set of moments of $\rho(t)$. For
example, if $c$ is an observable, then we can define the estimate
moments
\begin{eqnarray*}
\langle c \rangle (t)&\equiv&\textrm{Tr}[\rho(t) c]\\
\langle \Delta c^2 \rangle (t)&\equiv&\textrm{Tr}[\rho(t) (c-\langle c \rangle)^2]\\
\langle \theta \rangle (t) &\equiv& \int p(\theta,t) \theta d\theta\\
\langle \Delta\theta^2 \rangle (t)&\equiv& \int p(\theta,t)
(\theta-\langle \theta \rangle )^2 d\theta
\end{eqnarray*}
and derive their update rules from the full update rules,
resulting in a set of $y(t)$-dependent differential equations.  If
those differential equations are closed, then this reduced
description is adequate for the parameter estimation task at hand.
This situation (with closure and Gaussian distributions) is to be
expected when the system is approximately linear.

\subsection{Continuously measured spin system}\label{Section::QPESpins}

This approach can be applied directly to the problem of
magnetometry considered in this paper. The problem can be
summarized by the situation illustrated in \reffig{Schematic}: a
spin ensemble of possibly unknown number is initially polarized
along the x-axis (e.g., via optical pumping), an unknown possibly
fluctuating scalar magnetic field $b$ directed along the y-axis
causes the spins to then rotate within the x-z plane, and the
z-component of the collective spin is measured continuously.  The
measurement can, for example, be implemented as shown, where we
observe the difference photocurrent, $y(t)$, in a polarimeter
which measures the Faraday rotation of a linearly polarized far
off resonant probe beam travelling along z \cite{Geremia2003b,
Jessen2003, Deutsch2003}. The goal is to optimally estimate $b(t)$
via the measurement record and unbiased prior information. If a
control field $u(t)$ is included, as it will be eventually, the
total field is represented by $h(t)=b(t)+u(t)$.

In terms of our previous discussion, we have here the observable
$c= \sqrt{M}J_z$, where $M$ is the measurement rate (defined in
terms of probe beam parameters), and the parameter $\theta = b$.
When $b$ is known, our state estimate evolves by the stochastic
master equation \cite{Thomsen2002}
\begin{eqnarray}
d\rho_b(t)=-i[H(b),\rho_b(t)]dt+\mathcal{D}[\sqrt{M}J_z]\rho_b(t)dt\nonumber\\
+\sqrt{\eta}\mathcal{H}[\sqrt{M}J_z]\left(2 \sqrt{M\eta}[y(t)
dt-\langle J_z\rangle_b dt]\right)\rho_b(t)\label{Equation::SME}
\end{eqnarray}
where $H(b)=\gamma J_y b$, $\gamma$ is the gyromagnetic ratio, and
\begin{eqnarray*}
\mathcal{D}[c]\rho&\equiv& c \rho
c^{\dagger}-(c^{\dagger}c\rho+\rho c^{\dagger}c)/2\\
\mathcal{H}[c]\rho&\equiv&c\rho+\rho
c^{\dagger}-\textrm{Tr}[(c+c^{\dagger})\rho]\rho
\end{eqnarray*}
The stochastic quantity $2 \sqrt{M\eta}[y(t) dt-\langle
J_z\rangle_b (t) dt]\equiv d\bar{W}(t)$ is a Wiener increment
(Gaussian white noise with variance $dt$) by the optimality of the
filter. The sensitivity of the photodetection per
$\sqrt{\textrm{Hz}}$ is represented by $1/2\sqrt{M\eta}$, where
the quantity $\eta$ represents the quantum efficiency of the
detection.  If $\eta = 0$, we are essentially ignoring the
measurement result and the conditional SME becomes a deterministic
unconditional master equation.  If $\eta=1$, the detectors are
maximally efficient. Note that our initial state
$\rho(0)=\rho_b(0)$ is made equal to a coherent state (polarized
in x) and is representative of our prior information.

It can be shown that the unnormalized probability $\bar{p}(b,t)$
evolves according to \cite{Verstraete2001}
\begin{eqnarray}
d\bar{p}(b,t)=4M\eta\langle J_z\rangle_b (t) \bar{p}(b,t)y(t)
dt\label{Equation::ProbabilityUpdate}
\end{eqnarray}
The evolution \refeqns{SME}{ProbabilityUpdate} together with
\refeqn{FullEstimate} solve the problem completely, albeit in a
computationally expensive way. Clearly, for large ensembles it
would be advantageous to reduce the problem to a simpler
description.

If we consider only the estimate moments $\langle J_z \rangle(t)$,
$\langle \Delta J_z^2 \rangle(t)$, $\langle b \rangle(t)$, and
$\langle \Delta b^2 \rangle(t)$ and derive their evolution with
the above rules, it can be shown that the filtering equations for
those variables are closed under certain approximations.  First,
the spin number $J$ must be large enough that the distributions
for $J_y$ and $J_z$ are approximately Gaussian for an x-polarized
coherent state. Second, we only consider times $t\ll 1/M$ because
the total spin becomes damped by the measurement at times
comparable to the inverse of the measurement rate.

Although this approach is rigorous and fail-safe, the resulting
filtering equations for the moments can be arrived at in a more
direct manner as discussed in \refapx{SimplifiedRepresentation}.
Essentially, the full quantum mechanical mapping from $h(t)$ to
$y(t)$ is equivalent to the mapping derived from a model which
appears classical, and assumes an actual, but random, value for
the $z$ component of spin. This correspondence generally holds for
a stochastic master equation corresponding to an arbitrary linear
quantum mechanical system with continuous measurement of
observables that are linear combinations of the canonical
variables \cite{Doherty2003}.

From this point on we will only consider the simplified Gaussian
representation (used in the next section) since it allows us to
apply established techniques from estimation and control theory.
The replacement of the quantum mechanical model with a classical
noise model is discussed more fully in the appendix. Throughout
this treatment, we keep in mind the constraints that the original
model imposed. As before, we assume $J$ is large enough to
maintain the Gaussian approximation and that time is small
compared to the measurement induced damping rate, $t\ll 1/M$.
Also, the description of our original problem demands that
$\langle \Delta J_z^2 \rangle(0)=J/2$ for a coherent state
\footnote{We assume throughout the paper that we have a system of
$N$ spin-$1/2$ particles, so for a polarized state along $x$,
$\langle J_x \rangle = J = N/2$ and $\sigma_{z0}=\langle \Delta
J_z^2 \rangle(0)=J/2=N/4$. This is an arbitrary choice and our
results are independent of any constituent spin value, apart from
defining these moments. In \cite{Geremia2003b}, for example, we
work with an ensemble of Cs atoms, each atom in a ground state of
spin-$4$. \label{Convention}}. Hence our prior information for the
initial value of the spin component will always be dictated by the
structure of Hilbert space.

\section{Optimal estimation and control}\label{Section::OptimalEstimationAndControl}

We now describe the dynamics of the simplified representation.
Given a linear state-space model (L), a quadratic performance
criterion (Q) and Gaussian noise (G), we show how to apply
standard LQG analysis to optimize the estimation and control
performance \cite{Jacobs1996}.

The system state we are trying to estimate is represented by
\begin{flushleft}
\textbf{State}
\begin{eqnarray}
\vec{x}(t)&\equiv&\begin{bmatrix} z(t) \\
b(t)\label{Equation::State}
\end{bmatrix}
\end{eqnarray}
\end{flushleft}
where $z(t)$ represents the small z-component of the collective
angular momentum and $b(t)$ is a scalar field along the y axis.

Our best guess of $\vec{x}(t)$, as we filter the measurement
record, will be denoted
\begin{flushleft}
\textbf{Estimate}
\begin{eqnarray}
\vec{m}(t)&\equiv&\begin{bmatrix} \tilde{z}(t) \\
\tilde{b}(t)\end{bmatrix}\label{Equation::Estimate}
\end{eqnarray}
\end{flushleft}
As stated in the appendix, we implicitly make the associations:
$\tilde{z}(t)  =  \langle J_z \rangle (t)=\textrm{Tr}[\rho(t)
J_z]$ and $\tilde{b}(t) = \int p(b,t) b \,db$, although no further
mention of $\rho(t)$ or $p(b,t)$ will be made.

We assume the measurement induced damping of $J$ to be negligible
for short times ($J \exp[-M t/2]\approx J$ if $t \ll 1/M$) and
approximate the dynamics as
\begin{flushleft}
\textbf{Dynamics}
\begin{eqnarray}
d\vec{x}(t)&=&\textbf{A}\vec{x}(t)dt+\textbf{B}u(t)
dt+\begin{bmatrix} 0 \\
\sqrt{\sigma_{bF}}\end{bmatrix}dW_1 \label{Equation::Dynamics}\\
\textbf{A}&\equiv&\begin{bmatrix} 0 & \gamma J \\ 0 &
-\gamma_b\end{bmatrix}\nonumber\\
\textbf{B}&\equiv&\begin{bmatrix} \gamma J\\
0\end{bmatrix}\nonumber\\
\mathbf{\Sigma_0}&\equiv&\begin{bmatrix} \sigma_{z0} & 0 \\ 0 &
\sigma_{b0}\end{bmatrix}\nonumber\\
\mathbf{\Sigma_1}&\equiv&\begin{bmatrix} 0 & 0 \\ 0 &
\sigma_{bF}\end{bmatrix}\nonumber
\end{eqnarray}
\end{flushleft}
where the initial value $\vec{x}(0)$ for each trial is drawn
randomly from a Gaussian distribution of mean zero and covariance
matrix $\mathbf{\Sigma}_0$.  The initial field variance
$\sigma_{b0}$ is considered to be due to classical uncertainty,
whereas the initial spin variance $\sigma_{z0}$ is inherently
non-zero due to the original quantum state description.
Specifically, we impose $\sigma_{z0}=\langle \Delta J_z ^2 \rangle
(0)$. The Wiener increment $dW_1(t)$ has a Gaussian distribution
with mean zero and variance $dt$. $\mathbf{\Sigma_1}$ represents
the covariance matrix of the last vector in \refeqn{Dynamics}.

We have given ourselves a magnetic field control input, $u(t)$,
along the same axis, y, of the field to be measured, $b(t)$. We
have allowed $b(t)$ to fluctuate via a damped diffusion
(Ornstein-Uhlenbeck) process \cite{Gardiner2002}
\begin{eqnarray}
db(t)&=&-\gamma_b b(t)\,dt+\sqrt{\sigma_{bF}}\,dW_1
\label{Equation::bFluctuations}
\end{eqnarray}
The $b(t)$ fluctuations are represented in this particular way
because Gaussian noise processes are amenable to LQG analysis. The
variance of the field at any particular time is given by the
expectation $\sigma_{bFree}\equiv\textrm{E}[b(t)^2]=\sigma_{bF}/2
\gamma_b$. (Throughout the paper we use the notation
$\textrm{E}[x(t)]$ to represent the average of the generally
stochastic variable $x(t)$ at the same point in time, over many
trajectories.) The bandwidth of the field is determined by the
frequency $\gamma_b$ alone. When considering the measurement of
fluctuating fields, a valid choice of prior might be
$\sigma_{b0}=\sigma_{bFree}$, but we choose to let $\sigma_{b0}$
remain independent.  For constant fields, we set
$\sigma_{bFree}=0$, but $\sigma_{b0}\neq 0$.

Note that only the small angle limit of the spin motion is
considered. Otherwise we would have to consider different
components of the spin vector rotating into each other.  The small
angle approximation would be invalid if a field caused the spins
to rotate excessively, but using adequate control ensures this
will not happen. Hence, we use control for essentially two reasons
in this paper: first to keep our small angle approximation valid,
and, second, to make our estimation process robust to our
ignorance of $J$.  The latter point will be discussed in
\refsec{RobustPerformance}.

Our measurement of $z$ is described by the process
\begin{flushleft}
\textbf{Measurement}
\begin{eqnarray}
y(t) dt&=&\textbf{C}\vec{x}(t)dt + \sqrt{\sigma_M}dW_2(t)\label{Equation::Measurement}\\
\textbf{C}&\equiv&\begin{bmatrix} 1 & 0\end{bmatrix}\nonumber\\
\mathbf{\Sigma_2}&\equiv&\sigma_M\equiv 1/4M\eta\nonumber
\end{eqnarray}
\end{flushleft}
where the measurement shot-noise is represented by the Wiener
increment $dW_2(t)$ of variance $dt$.  Again, $\sqrt{\sigma_M}$
represents the sensitivity of the measurement, $M$ is the
measurement rate (with unspecified physical definition in terms of
probe parameters), and $\eta$ is the quantum efficiency of the
measurement. The increments $dW_1$ and $dW_2$ are uncorrelated.

Following \cite{Jacobs1996}, the optimal estimator for mapping
$y(t)$ to $\vec{m}(t)$ takes the form
\begin{flushleft}
\textbf{Estimator}
\begin{eqnarray}
d\vec{m}(t)&=&\textbf{A}\vec{m}(t)dt+\textbf{B}u(t) dt
\nonumber\\
&&+\textbf{K}_O(t)[y(t)-\textbf{C}\vec{m}(t)]dt \label{Equation::KalmanFilter}\\
\vec{m}(0)&=&\begin{bmatrix} 0\\ 0 \end{bmatrix} \nonumber\\
\textbf{K}_O(t)&\equiv&\mathbf{\Sigma}(t)\textbf{C}^T
\mathbf{\Sigma}_{2}^{-1}\nonumber\\
\frac{d\mathbf{\Sigma}(t)}{dt}&=&\mathbf{\Sigma}_1+\textbf{A}\mathbf{\Sigma}(t)+ \mathbf{\Sigma}(t)\textbf{A}^{T}\nonumber\\
&&-\mathbf{\Sigma}(t)\textbf{C}^{T}\mathbf{\Sigma}_{2}^{-1}\textbf{C}\mathbf{\Sigma}(t) \label{Equation::Riccati}\\
\mathbf{\Sigma}(t)&\equiv&\begin{bmatrix} \sigma_{zR}(t) & \sigma_{cR}(t)\\ \sigma_{cR}(t) & \sigma_{bR}(t)\end{bmatrix}\label{Equation::Sigma}\\
\mathbf{\Sigma}(0)&=&\mathbf{\Sigma_0}\equiv\begin{bmatrix}
\sigma_{z0} & 0
\\ 0 & \sigma_{b0}\end{bmatrix}\label{Equation::Priors}
\end{eqnarray}
\end{flushleft}

\refeqn{KalmanFilter} is the Kalman filter which depends on the
solution of the matrix Riccati \refeqn{Riccati}. The Riccati
equation gives the optimal observation gain $\textbf{K}_O(t)$ for
the filter. The estimator is designed to minimize the average
quadratic estimation error for each variable:
$\textrm{E}[(z(t)-\tilde{z}(t))^2]$ and
$\textrm{E}[(b(t)-\tilde{b}(t))^2]$. If the model is correct, and
we assume the observer chooses his prior information
$\mathbf{\Sigma}(0)$ to match the actual variance of the initial
data $\mathbf{\Sigma_0}$, then we have the self-consistent result:
\begin{eqnarray*}
\sigma_{zE}(t)&\equiv&\textrm{E}[(z(t)-\tilde{z}(t))^2]=\sigma_{zR}(t)\\
\sigma_{bE}(t)&\equiv&\textrm{E}[(b(t)-\tilde{b}(t))^2]=\sigma_{bR}(t)
\end{eqnarray*}
Hence, the Riccati equation solution represents both the observer
gain \emph{and} the expected performance of an optimal filter
using that same gain.

Now consider the control problem, which is in many respects dual
to the estimation problem.  We would like to design a controller
to map $y(t)$ to $u(t)$ in a manner that minimizes the quadratic
cost function
\begin{flushleft}
\textbf{Minimized Cost}
\begin{eqnarray}
H&=&\int_0^T \left[\vec{x}^T(t) \textbf{P} \vec{x}(t) + u(t)
\textbf{Q} u(t)\right] \,dt\nonumber \\
&&+\vec{x}^T(T) \textbf{P}_1 \vec{x}(T) \label{Equation::ControllerCost}\\
\mathbf{P}&\equiv&\begin{bmatrix} p & 0
\\ 0 & 0\end{bmatrix}\nonumber\\
\mathbf{Q}&\equiv&q\nonumber
\end{eqnarray}
\end{flushleft}
where $\textbf{P}_1$ is the end-point cost.  Only the ratio $p/q$
ever appears, of course, so we define the parameter
$\lambda\equiv\sqrt{p/q}$ and use it to represent the cost of
control. By setting $\lambda\rightarrow \infty$, as we often
choose to do in the subsequent analysis to simplify results, we
are putting no cost on our control output. This is unrealistic
because, for example, making $\lambda$ arbitrarily large implies
that we can apply transfer functions with finite gain at
arbitrarily high frequencies, which is not experimentally
possible. Despite this, we will often consider the limit
$\lambda\rightarrow \infty$ to set bounds on achievable estimation
and control performance. The optimal controller for minimizing
\refeqn{ControllerCost} is
\begin{flushleft}
\textbf{Controller}
\begin{eqnarray}
u(t)&=&-\textbf{K}_C(t) \vec{m}(t)\\
\textbf{K}_C(t)&\equiv&\textbf{Q}^{-1}\textbf{B}^T \textbf{V}(T-t)\nonumber\\
\frac{d\textbf{V}(T)}{dT}&=&\textbf{P}+\textbf{A}^{T}\textbf{V}(T)+\textbf{V}(T)\textbf{A} \nonumber\\
&&-\textbf{V}(T)\textbf{B}\textbf{Q}^{-1}\textbf{B}^T\textbf{V}(T)\label{Equation::VRiccati}\\
\textbf{V}(T=0)&\equiv&\textbf{P}_1\nonumber
\end{eqnarray}
\end{flushleft}
Here $\textbf{V}(T)$ is solved in reverse time $T$, which can be
interpreted as the time left to go until the stopping point. Thus
if $T\rightarrow\infty$, then we only need to use the steady state
of the $\textbf{V}$ Riccati \refeqn{VRiccati} to give the steady
state controller gain $\textbf{K}_C$ for all times. In this case,
we can ignore the (reverse) initial condition $P_1$ because the
controller is not designed to stop.  Henceforth, we will make
$\textbf{K}_C$ equal to this constant steady state value, such
that the only time varying coefficients will come from
$\textbf{K}_O(t)$.

In principle, the above results give the entire solution to the
ideal estimation and control problem.  However, in the non-ideal
case where our knowledge of the system is incomplete, e.g. $J$ is
unknown, our estimation performance will suffer.  Notation is now
introduced which produces trivial results in the ideal case, but
is helpful otherwise.  Our goal is to collect the above equations
into a single structure which can be used to solve the non-ideal
problem. We define the \emph{total} state of the system and
estimator as
\begin{flushleft}
\textbf{Total State}
\begin{eqnarray}
 \vec{\theta}(t)&\equiv&\begin{bmatrix} \vec{x}(t) \\
\vec{m}(t)
\end{bmatrix}=\begin{bmatrix} z(t) \\ b(t) \\ \tilde{z}(t) \\ \tilde{b}(t)
\end{bmatrix}
\end{eqnarray}
\end{flushleft}

Consider the general case where the observer assumes the plant
contains spin $J'$, which may or may not be equal to the actual
$J$. All design elements depending on $J'$ instead of $J$ are now
labelled with a prime.  Then it can be shown that the total state
dynamics from the above estimator-controller architecture are a
time-dependent Ornstein-Uhlenbeck process
\begin{flushleft}
\textbf{Total State Dynamics}
\begin{eqnarray}
d\vec{\theta}(t) &=& \mathbf{\alpha}(t) \vec{\theta}(t) dt + \mathbf{\beta}(t) d\vec{W}(t)\\
\mathbf{\alpha}(t)&\equiv&\begin{bmatrix} \textbf{A} & -\textbf{B}\textbf{K}'_C \\
\textbf{K}'_O(t) \textbf{C} &
\textbf{A}'-\textbf{B}'\textbf{K}'_C-\textbf{K}'_O(t) \textbf{C}
\end{bmatrix}\nonumber\\
\mathbf{\beta}(t)&\equiv&\begin{bmatrix}0 & 0 & 0 & 0 \\
0 & \sqrt{\sigma_{bF}} & 0 & 0 \\
0 & 0 & \sqrt{\sigma_M} K'_{O1}(t) & 0 \\
0 & 0 & \sqrt{\sigma_M} K'_{O2}(t)  & 0
\end{bmatrix}\nonumber
\end{eqnarray}
\end{flushleft}
where  the covariance matrix of $d\vec{W}$ is $dt$ times the
identity. Now the quantity of interest is the following covariance
matrix
\begin{flushleft}
\textbf{Total State Covariance}
\begin{eqnarray}
\mathbf{\Theta}(t)&\equiv& \textrm{E}[\vec{\theta}(t)\vec{\theta}^T(t)]\nonumber\\
&\equiv&\begin{bmatrix}\sigma_{zz} & \sigma_{zb} & \sigma_{z\tilde{z}} & \sigma_{z\tilde{b}} \\
\sigma_{zb} & \sigma_{bb} & \sigma_{b\tilde{z}} & \sigma_{b\tilde{b}} \\
\sigma_{z\tilde{z}} & \sigma_{b\tilde{z}} & \sigma_{\tilde{z}\tilde{z}} & \sigma_{\tilde{z}\tilde{b}} \\
\sigma_{z\tilde{b}} & \sigma_{b\tilde{b}} &
\sigma_{\tilde{z}\tilde{b}} &
\sigma_{\tilde{b}\tilde{b}} \end{bmatrix}\\
\sigma_{zz}&\equiv&\textrm{E}[z(t)^2]\nonumber\\
\sigma_{zb}&\equiv&\textrm{E}[z(t)b(t)]\nonumber\\
\vdots &\equiv&\vdots\nonumber
\end{eqnarray}
\end{flushleft}
It can be shown that this total covariance matrix obeys the
deterministic equations of motion
\begin{flushleft}
\textbf{Total State Covariance Dynamics}
\begin{eqnarray}
\frac{d\mathbf{\Theta}(t)}{dt}&=&\mathbf{\alpha}(t) \mathbf{\Theta}(t)+\mathbf{\Theta}(t) \mathbf{\alpha}^T(t) + \mathbf{\beta}(t) \mathbf{\beta}^T(t)\label{Equation::TotalDifferential}\\
\mathbf{\Theta}(t)&=&\exp\left[-\int_{0}^{t}\mathbf{\alpha}(t')
dt'\right]\mathbf{\Theta}_0\exp\left[-\int_{0}^{t}\mathbf{\alpha}^T(t')
dt'\right]\nonumber\\
&&+\int_0^tdt'\exp\left[-\int_{t'}^{t}\mathbf{\alpha}(s)
ds\right]\mathbf{\beta}(t')\mathbf{\beta}^T(t')\nonumber\\
&&\quad\times\exp\left[-\int_{t'}^{t}\mathbf{\alpha}^T(s) ds\right] \label{Equation::TotalIntegral}\\
\mathbf{\Theta}_0&=&\begin{bmatrix}\sigma_{z0} & 0 & 0 & 0 \\
0 & \sigma_{b0} & 0 & 0 \\
0 & 0 & 0 & 0 \\
0 & 0 & 0  & 0 \nonumber
\end{bmatrix}
\end{eqnarray}
\end{flushleft}
\refeqn{TotalIntegral} is the matrix form of the standard
integrating factor solution for time-dependent scalar ordinary
differential equations \cite{Gardiner2002}. Whether we solve this
problem numerically or analytically, the solution provides the
quantity that we ultimately care about
\begin{flushleft}
\textbf{Average Magnetometry Error}
\begin{eqnarray}
\sigma_{bE}(t)&\equiv& \textrm{E}[(\tilde{b}(t)-b(t))^2]\nonumber\\
&=&\textrm{E}[b^2(t)]+\textrm{E}[\tilde{b}^2(t)]-2\textrm{E}[b(t)\tilde{b}(t)]\nonumber\\
&=&\sigma_{bb}(t)+\sigma_{\tilde{b}\tilde{b}}(t)-2\sigma_{b\tilde{b}}(t)\label{Equation::AveMagnetometryError}
\end{eqnarray}
\end{flushleft}

When all parameters are known (and $J'=J$), this total state
description is unnecessary because
$\sigma_{bE}(t)=\sigma_{bR}(t)$. This equality is by
\emph{design}. However, when the wrong parameters are assumed
(e.g., $J'\neq J$) the equality does not hold $\sigma_{bE}(t)\neq
\sigma_{bR}(t)$ and either \refeqn{TotalDifferential} or
\refeqn{TotalIntegral} must be used to find $\sigma_{bE}(t)$.
Before addressing this problem, we consider in detail the
performance in the ideal case, where all system parameters are
known by the observer, including $J$.

At this point, we have defined several variables.  For clarity,
let us review the meaning of several before continuing.  Inputs to
the problem include the field fluctuation strength $\sigma_{bF}$,
\refeqn{bFluctuations}, and the measurement sensitivity
$\sigma_M$, \refeqn{Measurement}. The prior information for the
field is labelled $\sigma_{b0}$, \refeqn{Priors}. The solution to
the Riccati equation is $\sigma_{bR}(t)$, \refeqn{Sigma}, and is
equal to the estimation variance $\sigma_{bE}(t)$,
\refeqn{AveMagnetometryError}, when the estimator model is
correct. In the next section, we additionally use $\sigma_{bS}$,
\refeqn{bSteady}, and $\sigma_{bT}(t)$, \refeqn{bTransient}, to
represent the steady state and transient values of
$\sigma_{bE}(t)$ respectively.

\section{Optimal performance: $J$ known}\label{Section::OptimalPerformance}

We start by observing qualitative characteristics of the
$b$-estimation dynamics. \reffig{bRiccati} shows the average
estimation performance, $\sigma_{bR}(t)$, as a function of time
for a realistic set of parameters. Notice that $\sigma_{bR}$ is
constant for small and large times, below $t_1$ and above $t_2$.
If $\sigma_{b0}$ is non-infinite then the curve is constant for
small times, as it takes some time to begin improving the estimate
from the prior. If $\sigma_{b0}$ is infinite, then $t_1=0$ and the
sloped transient portion extends towards infinity as $t\rightarrow
0$. At long times, $\sigma_{bR}$ will become constant again, but
only if the field is fluctuating ($\sigma_{bF}\neq 0$ and
$\gamma_b\neq 0$). The performance saturates because one can track
a field only so well if the field is changing and the
signal-to-noise ratio is finite. If the field to be tracked is
constant, then $t_2=\infty$ and the sloped portion of the curve
extends to zero as $t\rightarrow \infty$ (given the approximations
discussed in \refsec{QPESpins}). After the point where the
performance saturates ($t\gg t_2$), all of the observer and
control gains have become time-independent and the filter can be
described by a transfer function.

\begin{figure}
\includegraphics[width=3.25in]{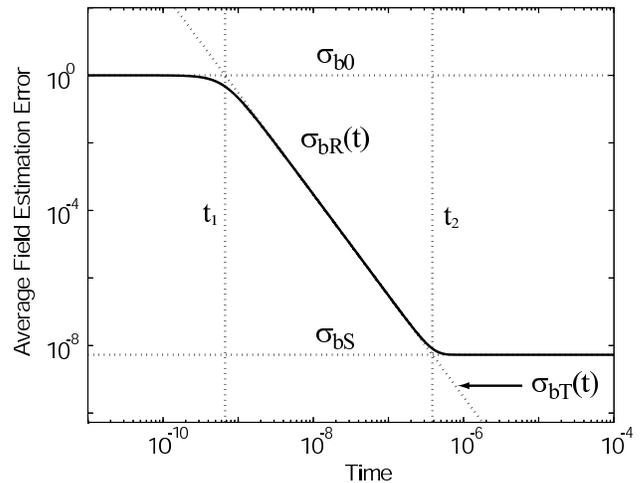} \caption{
The Riccati equation solution gives the ideal field estimation
performance. The parameters used here are $J=10^6$,
$\sigma_{z0}=J/2$ (for ensemble of spin-$1/2$'s), $\gamma=10^6$,
$M=10^4$, $\sigma_{b0}=\sigma_{bFree}=1$. The solution starts at
the free field fluctuation variance and saturates at
$\sigma_{bS}$. The plot is not valid at times $t>
1/M$.}\label{Figure::bRiccati}
\end{figure}

However, as will be shown, applying only this steady state
transfer function is non-optimal in the transient regime ($t_1\ll
t\ll t_2$), because the time dependence of the gains is clearly
crucial for optimal transient performance.

\subsection{Steady state performance}

We start by examining the steady state performance of the filter.
At large enough times (where we have yet to define large enough),
$\textbf{K}_O$ becomes constant and if we set $T\rightarrow
\infty$ (ignoring the end point cost), then $\textbf{K}_C$ is
always constant. Setting $d\mathbf{\Sigma}/dt=0$ and
$d\textbf{V}/dt=0$ we find:
\begin{eqnarray*}
\textbf{K}_O(t)&\rightarrow&
\begin{bmatrix}
\sqrt{2 \gamma J
\sqrt{\frac{\sigma_{bF}}{\sigma_M}}+\gamma_b^2}-\gamma_b\\
\sqrt{\frac{\sigma_{bF}}{\sigma_M}}-\frac{\gamma_b}{\gamma
J}(\sqrt{2 \gamma J
\sqrt{\frac{\sigma_{bF}}{\sigma_M}}+\gamma_b^2}-\gamma_b)
\end{bmatrix}\\
\textbf{K}_C(t)&\rightarrow&\begin{bmatrix} \lambda &
1/(1+\frac{\gamma_b}{\gamma J \lambda})
\end{bmatrix}
\end{eqnarray*}
where $\lambda= \sqrt{\frac{p}{q}}$.

Now assuming the gains to be constant, we can derive the three
relevant transfer functions from $y(t)$ to $\vec{m}(t)$
($\tilde{z}$ and $\tilde{b}$) and $u$. We proceed as follows.
First, we express the estimates in terms of only themselves and
the photocurrent
\begin{eqnarray*}
\frac{d\vec{m}(t)}{dt}&=&\textbf{A}\vec{m}(t)+\textbf{B}u(t) +
\textbf{K}_O(y(t)-\textbf{C}\vec{m}(t))\\
&=&\textbf{A}\vec{m}(t)+\textbf{B}(-\textbf{K}_C \vec{m}(t)) +
\textbf{K}_O(y(t)-\textbf{C}\vec{m}(t))\\
&=&(\textbf{A}-\textbf{B}\textbf{K}_C-\textbf{K}_O\textbf{C})\vec{m}(t)+
\textbf{K}_O y(t)
\end{eqnarray*}
To get the transfer functions, we take the Laplace transform of
the entire equation, use differential transform rules to give $s$
factors (where $s= j\omega$, $j=\sqrt{-1}$), ignore initial
condition factors, and rearrange terms. However, this process only
gives meaningful transfer functions if the coefficients
$\textbf{K}_O$ and $\textbf{K}_C$ are constant. Following this
procedure, we have
\begin{eqnarray*}
\vec{m}(s)&=&(s\textbf{I}-\textbf{A}+\textbf{B}\textbf{K}_C+\textbf{K}_O\textbf{C})^{-1}\textbf{K}_O
y(s)\\
&=& \vec{G}_m(s)y(s)\\
u(s)&=&-\textbf{K}_C \vec{m}(s)\\
&=&-\textbf{K}_C(s-\textbf{A}+\textbf{B}\textbf{K}_C+\textbf{K}_O\textbf{C})^{-1}\textbf{K}_O
y(s)\\
&=& G_u(s) y(s)
\end{eqnarray*}
where
\begin{eqnarray*}
\vec{G}_m(s)=\begin{bmatrix} G_z(s)\\ G_b(s)
\end{bmatrix}
\end{eqnarray*}

The three transfer functions ($G_z(s)$, $G_b(s)$, and $G_u(s)$)
serve three different tasks.  If estimation is the concern, then
$G_b(s)$ will perform optimally in steady state. Notice that,
while the Riccati solution is the same with and without control
($\textbf{K}_C$ non-zero or zero), this transfer function is not
the same in the two cases. So, even though the transfer functions
are different, they give the same steady state performance.

Let us now consider the controller transfer function $G_u(s)$ in
more detail. We find the controller to be of the form
\begin{equation}
G_{u}(s)=G_{u,DC}\frac{1+s/\omega_H}{1+(1+s/\omega_Q)s/\omega_L} \\
\end{equation}
Here each frequency $\omega$ represents a transition in the Bode
plot of \reffig{Bode}.  A similar controller transfer function is
derived via a different method in \refapx{RobustFrequencySpace}.

If we are not constrained experimentally, we can make the
approximations
$\lambda^2\gg\sqrt{\sqrt{\sigma_{bF}/\sigma_M}/2\gamma J}$ and
$\gamma J\gg\gamma_b^2\sqrt{\sigma_{M}/\sigma_{bF}}$ giving
\begin{eqnarray*}
G_u(s)&\rightarrow&G_{u,DC}\frac{1+s/\omega_H}{1+s/\omega_L} \\
\omega_L &\rightarrow& \gamma_b\\
\omega_H&\rightarrow&\sqrt{\frac{\gamma J}{2}\sqrt{\frac{\sigma_{bF}}{\sigma_{M}}}}\\
\omega_C&\rightarrow&\sqrt{2\gamma
J\sqrt{\frac{\sigma_{bF}}{\sigma_{M}}}}=2 \omega_H\\
\omega_Q&\rightarrow&\lambda \gamma J\\
G_{u,DC}&\rightarrow&-\frac{1}{\gamma_b}\sqrt{\frac{\sigma_{bF}}{\sigma_{M}}} \\
G_{u,AC}&\rightarrow&G_{u,DC}\frac{\omega_L}{\omega_H}=-\sqrt{\frac{2}{\gamma
J}\sqrt{\frac{\sigma_{bF}}{\sigma_{M}}}}
\end{eqnarray*}
where $G_{u,AC}$ is the gain at high frequencies
($\omega>\omega_H$) and we find the closing frequency $\omega_C$
from the condition $|P_z(j\omega_C)G_u(j\omega_C)|=1$, with the
plant transfer function being the normal integrator $P_z(s)=\gamma
J/s$. Notice that the controller closes in the very beginning of
the flat high frequency region (hence with adequate phase margin)
because $\omega_C=2 \omega_H$.

\begin{figure}
\includegraphics[width=3.25in]{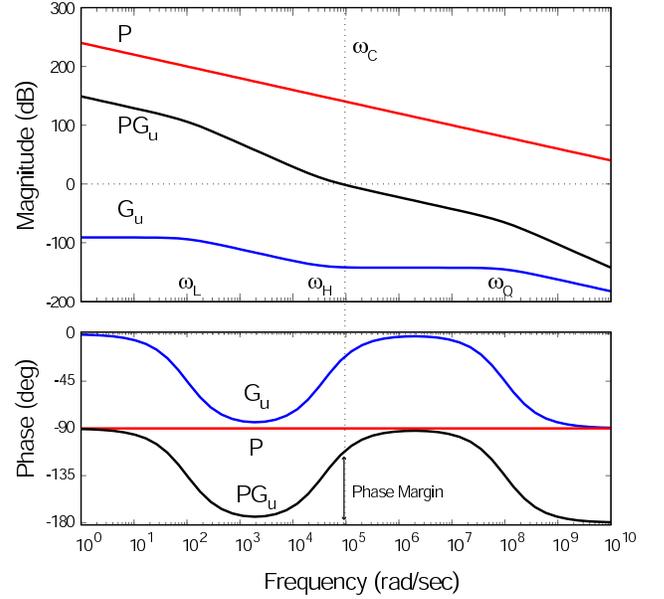} \caption{
The Bode plot of $G_u(s)$, the transfer function of the filter in
steady state, for a typical parameter regime. Notice that the
controller closes the plant with adequate phase margin to avoid
closed-loop instability. At high frequencies the controller rolls
off at $\omega_Q$ if $\lambda\neq \infty$.}\label{Figure::Bode}
\end{figure}

Finally, consider the steady state estimation performance. These
are the same with and without control (hence
$\lambda$-independent) and, under the simplifying assumption
$\gamma J\gg\gamma_b^2\sqrt{\sigma_{M}/\sigma_{bF}}$, are given by
\begin{eqnarray}
\sigma_{zR}(t)&\rightarrow& \sqrt{2 \gamma J} \sigma_{M}^{3/4}\sigma_{bF}^{1/4}\equiv \sigma_{zS}\label{Equation::zSteady}\\
\sigma_{bR}(t)&\rightarrow& \sqrt{\frac{2}{\gamma
J}}\sigma_{bF}^{3/4}\sigma_M^{1/4}\equiv
\sigma_{bS}\label{Equation::bSteady}
\end{eqnarray}
If the estimator reaches steady state at $t \ll 1/M$, then the
above variance $\sigma_{zR}$ represents a limit to the amount of
spin-squeezing possible in the presence of fluctuating fields.

Also the $J$ scaling of the saturated field sensitivity
$\sigma_{bR}\propto J^{-1/2}$ is not nearly as strong as the $J$
scaling in the transient period $\sigma_{bR}\propto J^{-2}$. Next,
we demonstrate this latter result as we move from the steady state
analysis to calculating the estimation performance during the
transient period.

\subsection{Transient performance}

We now consider the transient performance of the ideal filter: how
quickly and how well the estimator-controller will \emph{lock}
onto the signal and achieve steady state performance. In many
control applications, the transient response is not of interest
because the time it takes to acquire the lock is negligible
compared to the long steady state period of the system. However,
in systems where the measurement induces continuous decay, this
transient period can be a significant portion of the total
lifetime of the experiment.

\begin{table*}[t*]
\caption{Field tracking error, $\sigma_{bR}(t)$, for different
initial variances of $b$ and $z$.}
\begin{tabular*}{\textwidth}{c@{\hspace{.5cm}}|@{\hspace{1cm}}c@{\extracolsep{\fill}}cc@{\hspace{.5cm}}} \hline\hline
                                 & $\sigma_{b0}$ = 0 & $\sigma_{b0}$
                                                                    & $\sigma_{b0}\rightarrow \infty$                       \\ \hline
\rule[-3mm]{0mm}{8mm}
  $\sigma_{z0}$ = 0              & $0$               & $3\sigma_{b0}\sigma_M \left(3\sigma_M+\gamma^2 J^2 \sigma_{b0}t^3\right)^{-1}$
                                                                    & $3\sigma_M  \left( \gamma^2 J^2 t^3 \right)^{-1}$                  \\
\rule[-3mm]{0mm}{8mm}
  $\sigma_{z0}$                  & $0$               & $\frac{12\sigma_{b0}\sigma_M (\sigma_M+\sigma_{z0}t)}{12\sigma_M^2+\gamma^2J^2\sigma_{b0}\sigma_{z0}t^4 +4\sigma_M(3\sigma_{z0} t +\gamma^2 J^2 t^3 \sigma_{b0})}$
                                                                    & $12\sigma_M(\sigma_M+\sigma_{z0}t) \left(\gamma^2 J^2 t^3(4\sigma_M+\sigma_{z0}t)\right)^{-1}$\\
\rule[-3mm]{0mm}{8mm}
  $\sigma_{z0}\rightarrow\infty$ & $0$               & $12\sigma_{b0}\sigma_M \left(12\sigma_M+\gamma^2 J^2 t^3 \sigma_{b0}\right)^{-1}$
                                                                    & $12\sigma_M \left(\gamma^2 J^2 t^3\right)^{-1}$                 \\ \hline\hline
\end{tabular*}\label{Table::FieldVariance}
\end{table*}

\begin{table*}[t*]
\caption{Spin tracking error, $\sigma_{zR}(t)$, for different
initial variances of $b$ and $z$.}
\begin{tabular*}{\textwidth}{c@{\hspace{.5cm}}|@{\extracolsep{\fill}}ccc@{\hspace{.5cm}}} \hline\hline
                & $\sigma_{b0}=0$   & $\sigma_{b0}$ & $\sigma_{b0}\rightarrow\infty$                                    \\\hline
\rule[-3mm]{0mm}{8mm} $\sigma_{z0}=0$ & $0$               &
$3\gamma^2
                                        J^2\sigma_{b0} \sigma_M t^2 \left(3\sigma_M+\gamma^2 J^2\sigma_{b0}t^3 \right)^{-1}$
                                                    & $3\sigma_M t^{-1}$                                             \\
\rule[-3mm]{0mm}{8mm} $\sigma_{z0}$   &
$\sigma_M\sigma_{z0}\left(\sigma_M+
                    \sigma_{z0}t\right)^{-1}$
                                    &$\frac{4\sigma_M(\gamma^2J^2\sigma_{b0}
                                            \sigma_{z0}t^3+3\sigma_M(\sigma_{z0}+\gamma^2 J^2
                                            t^2\sigma_{b0}))}{12\sigma_M^2+\gamma^2J^2\sigma_{b0}\sigma_{z0}t^4
                                            +4\sigma_M(3\sigma_{z0} t +\gamma^2 J^2 t^3
                                            \sigma_{b0})}$
                                                    & $4\sigma_M(3\sigma_M+\sigma_{z0}t) \left( t(4\sigma_M+\sigma_{z0}t) \right)^{-1}$     \\
\rule[-3mm]{0mm}{8mm} $\sigma_{z0} \rightarrow \infty$
                & $\sigma_M t^{-1}$ & $4\sigma_M(3\sigma_M+\gamma^2J^2t^3\sigma_{b0}) \left(12\sigma_M
                                            t+\gamma^2 J^2 t^4\sigma_{b0}\right)^{-1}$
                                                    & $4\sigma_M t^{-1}$         \\ \hline\hline
\end{tabular*}\label{Table::SpinVariance}
\end{table*}

We will evaluate the transient performance of two different
filters. First, we look at the ideal dynamic version, with time
dependent observer gains derived from the Riccati equation. This
limits to a transfer function at long times when the gains have
become constant. Second, we numerically look at the case where the
same steady state transfer functions are used for the
\emph{entire} duration of the measurement. Because the gains are
not adjusted smoothly, the small time performance of this
estimator suffers. Of course, for long times the estimators are
equivalent.

\subsubsection{Dynamic estimation and control}

Now consider the transient response of $\mathbf{\Sigma}(t)$
(giving $\textbf{K}_O(t)$). We will continue to impose that
$\textbf{V}$ (thus $\textbf{K}_C$) is constant because we are not
interested in any particular stopping time.

The Riccati equation for $\mathbf{\Sigma}(t)$ (\refeqn{Riccati})
appears difficult to solve because it is non-linear.  Fortunately,
it can be reduced to a much simpler linear problem. See
\refapx{RiccatiSolution} for an outline of this method.

The solution to the fluctuating field problem ($\sigma_{bF}\neq 0$
and $\gamma_b\neq0$) is represented in \reffig{bRiccati}. This
solution is simply the constant field solution ($\sigma_{bF}= 0$
and $\gamma_b = 0$) smoothly saturating at the steady state value
of \refeqn{bSteady} at time $t_2$.  Thus, considering the long
time behavior of the constant field solution will tell us about
the transient behavior when measuring fluctuating fields. Because
the analytic form for the constant field solution is simple, we
consider only it and disregard the full analytic form of the
fluctuating field solution.

The analytic form of $\mathbf{\Sigma}(t)$ is highly instructive.
The general solutions to $\sigma_{bR}(t)$ and $\sigma_{zR}(t)$,
with arbitrary prior information $\sigma_{b0}$ and $\sigma_{z0}$,
are presented in the central entries of
\reftabs{FieldVariance}{SpinVariance} respectively. The other
entries of the tables represent the limits of these somewhat
complicated expressions as the prior information assumes extremely
large or small values.  Here, we notice several interesting
trade-offs.

First, the left hand column of \reftab{FieldVariance} is zero
because if a constant field is being measured, and we start with
complete knowledge of the field ($\sigma_{b0}=0$), then our job is
completed trivially. Now notice that if $\sigma_{b0}$ and
$\sigma_{z0}$ are both non-zero, then at long times we have the
lower right entry of \reftab{FieldVariance}
\begin{eqnarray}
\sigma_{bR}(t)=\frac{12\sigma_M}{\gamma^2 J^2
t^3}\equiv\sigma_{bT}(t)\label{Equation::bTransient}
\end{eqnarray}
This is the same result one gets when the estimation procedure is
simply to perform a least-squares line fit to the noisy
measurement curve for constant fields.  (Note that all of these
results are equivalent to the solutions of \cite{Geremia2003}, but
without $J$ damping.) If it were physically possible to ensure
$\sigma_{z0}=0$, then our estimation would improve by a factor of
four to the upper right result. However, quantum mechanics imposes
that this initial variance is non-zero (e.g., $\sigma_{z0}=J/2$
for a coherent state and less, but still non-zero, for a squeezed
state), and the upper right solution is unattainable.

Now consider the dual problem of spin estimation performance
$\sigma_{zR}(t)$ as represented in \reftab{SpinVariance}, where we
can make analogous trade-off observations. If there is no field
present, we set $\sigma_{b0}= 0$ and
\begin{eqnarray}
\sigma_{zR}(t)&=&\frac{\sigma_{z0}\sigma_M}{\sigma_M  + t
\sigma_{z0}}
\end{eqnarray}
When $\sigma_{zR}(t)$ is interpreted as the quantum variance
$\langle \Delta J_z^2\rangle (t)$, this is the ideal (non-damped)
conditional spin-squeezing result which is valid at $t \ll 1/M$,
before damping in $J$ begins to take effect \cite{Thomsen2002}. If
we consider the solution for $t\gg 1/JM$, we have the lower left
entry of \reftab{SpinVariance}, $\sigma_{zR}(t)=\sigma_M/t$.
However, if we must include constant field uncertainty in our
estimation, then our estimate becomes the lower right entry
$\sigma_{zR}(t)=4\sigma_M/t$ which is, again, a factor of four
worse.

If our task is field estimation, intrinsic quantum mechanical
uncertainty in $z$ limits our performance just as, if our task is
spin-squeezed state preparation, field uncertainty limits our
performance.

\subsubsection{Transfer function estimation and control}

Suppose that the controller did not have the capability to adjust
the gains in time as it tracked a fluctuating field.  One approach
would then be to apply the steady state transfer functions derived
above for the \emph{entire} measurement. While this approach
performs optimally in steady state, it approaches the steady state
in a non-optimal manner compared to the dynamic controller.
\reffig{TFTransient} demonstrates this poor transient performance
for tracking fluctuating fields of differing bandwidth. Notice
that the performance only begins to improve around the time that
the dynamic controller saturates.

Also notice that the transfer function $G_b(s)$ is dependent on
whether or not the state is being controlled, i.e. whether or not
$\lambda$ is zero. The performance shown in \reffig{TFTransient}
is for one particular value of $\lambda$, but others will give
different estimation performances for short times.  Still, all of
the transfer functions generated from any value of $\lambda$ will
limit to the same performance at long times.  Also, all of them
will perform poorly compared to the dynamic approach during the
transient time.

\section{Robust performance: $J$ unknown}\label{Section::RobustPerformance}

Until this point, we have assumed the observer has complete
knowledge of the system parameters, in particular the spin number
$J$.  We will now relax this assumption and consider the
possibility that, for each measurement, the collective spin $J$ is
drawn randomly from a particular distribution.  Although we will
be ignorant of a given $J$, we may still possess knowledge about
the distribution from which it is derived.  For example, we may be
certain that $J$ never assumes a value below a minimal value
$J_{min}$ or above a maximal value $J_{max}$.  This is a realistic
experimental situation, as it is unusual to have particularly long
tails on, for example, trapped atom number distributions.  We do
not explicitly consider the problem of $J$ fluctuating during an
individual measurement, although the subsequent analysis can
clearly be extended to this problem.

Given a $J$ distribution, one might imagine completely
re-optimizing the estimator-controller with the full distribution
information in mind. Our initial approach is more basic and in
line with robust control theory: we design our filter as before,
assuming a particular $J'$, then analyze how well this filter
performs on an ensemble with $J \neq J'$. With this information in
mind, we can decide if estimator-controllers built with $J'$ are
robust, with and without control, given the bounds on $J$. We will
find that, under certain conditions, using control makes our
estimates robust to uncertainty about the total spin number.

\begin{figure}[t]
\includegraphics[width=3.25in]{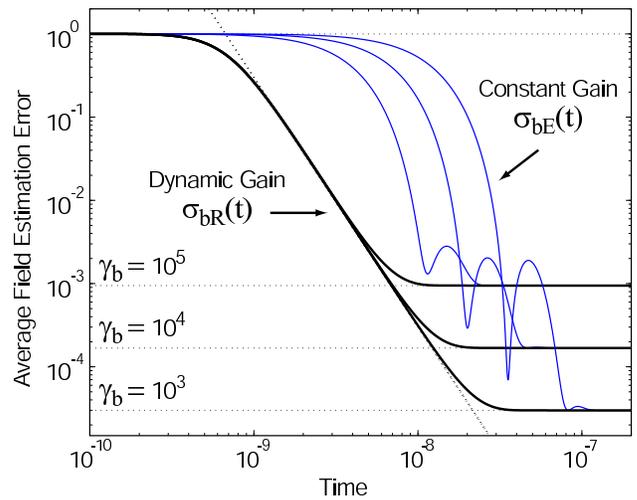} \caption{
Estimation performance for estimators based on the dynamic gain
solution of the Riccati equation, compared against estimators with
constant estimation gain.  The latter are the transfer function
limits of the former, hence they have the same long term
performance.  Three different bandwidth $b$ processes are
considered.}\label{Figure::TFTransient}
\end{figure}

The essential reason for this robustness is that when a control
field is applied to zero the measured signal, that control field
must be approximately equal to the field to be tracked.  Because
$J$ is basically an effective gain, variations in $J$ will affect
the performance, but not critically, so the error signal will
still be approximately zero.  If the applied signal is set to be
the estimate, then the tracking error must also be approximately
zero. (See \refapx{RobustFrequencySpace} for a robustness analysis
along these lines in frequency space. These methods were used in
\cite{Geremia2003b}, but neglect the transient behavior.)

Of course, this analysis assumes that we can apply fields with the
same precision that we measure them.  While the precision with
which we can apply a field is experimentally limited, we here
consider the ideal case of infinite precision.  In this admittedly
idealized problem, our estimation is limited by only the
measurement noise and our knowledge of $J$.

First, to motivate this problem, we describe how poorly our
estimator performs given ignorance about $J$ without control.

\subsection{Uncontrolled ignorance}

Let us consider the performance of our estimation procedure at
estimating constant fields when $J'\neq J$. In general, this
involves solving the complicated total covariance matrix
\refeqn{TotalIntegral}. However, in the long time limit ($t\gg
1/JM$) of estimating constant fields, the procedure amounts to
simply fitting a line to the noisy measurement with a
least-squares estimate.  Suppose we record an open-loop
measurement which appears as a noisy sloped line for small angles
of rotation due to the Larmor precession. Regardless of whether or
not we know $J$, we can measure the slope of that line and
estimate it to be $\tilde{m}$. If we knew $J$, we would know how
to extract the field from the slope correctly:
$\tilde{b}=\tilde{m}/\gamma J$. If we assumed the wrong spin
number, $J' \neq J$, we would get the non-optimal estimate:
$\tilde{b}'=\tilde{m}/\gamma J'=\tilde{b}J/J'$.

First assume that this is a \emph{systematic} error and $J$ is
unknown, but the same, on every trial.  We assume that the
constant field is drawn randomly from the $\sigma_{b0}$
distribution for every trial.  In this case, if we are wrong, then
we are always wrong by the same factor. It can be shown that the
error always saturates
\begin{eqnarray*}
\sigma_{bE}\rightarrow (1-f)^2\sigma_{b0}
\end{eqnarray*}
where $f = J/J'$. Of course, because this error is systematic, the
variance of the estimate does not saturate, only the error. This
problem is analogous to ignorance of the constant electronic gains
in the measurement and can also be calibrated away.

However, a significant problem arises when, on every trial, a
constant $b$ is drawn at random \emph{and} $J$ is drawn at random
from a distribution, so the error is no longer systematic. In this
case, we would not know whether to attribute the size of the
measured slope to the size of $J$ or to the size of $b$.  Given
the same $b$ every trial, all possible measurement curves fan out
over some angle due to the variation in $J$.  After measuring the
slope of an individual line to beyond this fan-out precision, it
makes no sense to continue measuring.

We should also point out procedures for estimating fields in
open-loop configuration, but \emph{without} the small angle
approximation. For constant large fields, we could observe many
cycles before the spin damped significantly.  By fitting the
amplitude and frequency independently, or computing the Fourier
transform, we could estimate the field somewhat independently of
$J$, which only determines the amplitude. However, the point here
is that $b$ might not be large enough to give many cycles before
the damping time or any other desired stopping time. In this case,
we could not independently fit the amplitude and frequency because
they appear as a product in the initial slope. Similar
considerations apply for the case of fluctuating $b$ and
fluctuating $J$. See \cite{Bretthorst1988}, for a complete
analysis of Bayesian spectrum analysis with free induction decay
examples.

Fortunately, using precise control can make the estimation process
relatively robust to such spin number fluctuations.

\subsection{Controlled ignorance: steady state performance}

We first analyze how the estimator designed with $J'$ performs on
a plant with $J$ at tracking fluctuating fields with and without
control. To determine this we calculate the steady state of
\refeqn{TotalDifferential}.

For the case of no control ($\lambda=0$), we simplify the
resulting expression by taking the same large $J'$ approximation
as before. This gives the steady state uncontrolled error
\begin{eqnarray*}
\sigma_{bE} &\rightarrow&
(1-f)^2\frac{\sigma_{bF}}{2\gamma_b}\\
&&=(1-f)^2\sigma_{bFree}
\end{eqnarray*}
where $f=J/J'$. Because the variance of the fluctuating $b$ is
$\sigma_{bFree}$, the uncontrolled estimation performs worse than
no estimation at all if $f>2$.

\begin{figure}[t]
\includegraphics[width=3.25in]{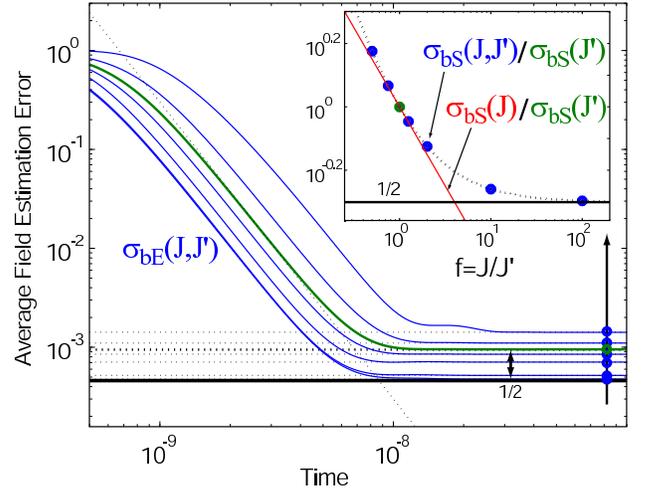} \caption{
Steady state estimation performance for estimator designed with
$J'=10^6$, and actual spin numbers: $J=J'\times [0.5,\, 0.75,\,
1,\, 1.25,\, 2,\, 10,\, 100]$. Other parameters: $\gamma=10^6$,
$M=10^4$, $\gamma_b=10^5$, $\sigma_{bFree}=1$ (fluctuating field),
$\lambda=0.1$ (this is large enough to satisfy large-$\lambda$
limits discussed in text).  The inset compares the normalized
robust estimation performance (curve) at a particular time, to the
ideal performance (line) when $J$ is known.
}\label{Figure::SteadyStateJp}
\end{figure}

\begin{figure}
\includegraphics[width=3.25in]{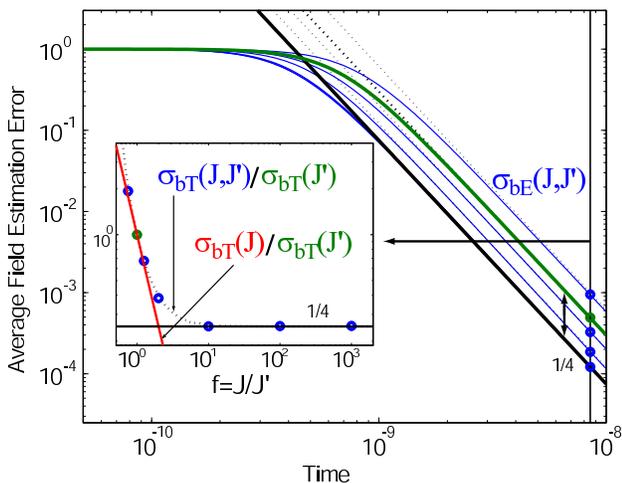} \caption{
Transient estimation performance for controller designed with
$J'=10^6$, and actual spin numbers: $J=J'\times [0.75,\, 1,\,
1.25,\, 2,\, 10,\, 100,\, 1000]$.  Other parameters:
$\gamma=10^6$, $M=10^4$, $\gamma_b=0$, $\sigma_{bFree}=0$
(constant field), $\lambda=1$.  Note that this behavior is valid
for $t<1/M=10^{-4}$. The inset compares the normalized robust
estimation performance (curve) at a particular time, to the ideal
performance (line) when $J$ is known.}\label{Figure::TransientJp}
\end{figure}

On the other hand, when we use precise control the performance
improves dramatically.  We again simplify the steady state
solution with the large $J'$ and $\lambda$ assumptions from
before, giving
\begin{eqnarray*}
\sigma_{bS}(J,J') &\rightarrow& \left(\frac{1+f}{2f}\right) \sqrt{\frac{2}{\gamma J'}}\sigma_{bF}^{3/4}\sigma_M^{1/4} \\
&=&\left(\frac{1+f}{2f}\right)\sigma_{bS}(J')
\end{eqnarray*}
where $\sigma_{bS}(J,J')$ is the steady state controlled error
when a plant with $J$ is controlled with a $J'$ controller and
$\sigma_{bS}(J')$ is the error when $J=J'$. One simple
interpretation of this result is that if we set $J'$ to be the
minimum of the $J$ distribution ($f>1$) then we never do worse
than $\sigma_{bS}(J')$ and we never do better than twice as well
($f\rightarrow\infty$). See \reffig{SteadyStateJp} for a
demonstration of this performance.

\subsection{Controlled ignorance: transient performance}

Now consider measuring constant fields with the wrong assumed
$J'$. Again, when control is not used, the error saturates at
\begin{eqnarray*}
\sigma_{bE}\rightarrow (1-f)^2\sigma_{b0}
\end{eqnarray*}
When control is used, the transient performance again improves
under certain conditions. The long time transient solution of
\refeqn{TotalDifferential} is difficult to manage analytically,
yet the behavior under certain limits is again simple.  For large
$\lambda$ and $J'$, and for $f>1/2$, we numerically find the
transient performance to be approximately
\begin{eqnarray}
\sigma_{bT}(J,J') &\rightarrow& \left(\frac{f^2+2}{4f^2-1}\right) \frac{12 \sigma_M}{\gamma^2 J'^2 t^3}\nonumber \\
&=&\left(\frac{f^2+2}{4f^2-1}\right)\sigma_{bT}(J')\label{Equation::Transientf}
\end{eqnarray}
where $\sigma_{bT}(J,J')$ is the transient controlled error when a
plant with $J$ is controlled with a $J'$ controller and
$\sigma_{bT}(J')$ is the error when $J=J'$.  See
\reffig{TransientJp} for a demonstration of this performance for
realistic parameters.  As $f\rightarrow\infty$ the $f$-dependent
pre-factor saturates at a value of $1/4$.  However, as
$f\rightarrow 1/2$ then the system takes longer to reach such a
simple asymptotic form, and the solution of \refeqn{Transientf}
becomes invalid.

Accordingly, one robust strategy would be the following.  Suppose
that the lower bound of the $J$-distribution was known and equal
to $J_{min}$.  Also assume that $\sigma_{bT}(J_{min})$ represents
an \emph{acceptable} level of performance. In this case, we could
simply design our estimator based on $J'=J_{min}$ and we would be
guaranteed at least the performance $\sigma_{bT}(J_{min})$ and at
best the performance $\sigma_{bT}(J_{min})/4$.

This approach would be suitable for experimental situations
because typical $J$ distributions are narrow: the difference
between $J_{min}$ and $J_{max}$ is rarely greater than an order of
magnitude. Thus, the overall sacrifice in performance between the
ideal case and the robust case would be small.  The estimation
performance still suffers because of our ignorance of $J$, but not
nearly as much as in the uncontrolled case.

\section{Conclusion}\label{Section::Conclusion}

The analysis of this paper contained several key steps which
should be emphasized. Our first goal was to outline the proper
approach to quantum parameter estimation. The second was to
demonstrate that reduced representations of the full filtering
problem are relevant and convenient because, if a simple
representation can be found, then existing classical estimation
and control methods can be readily applied. The characteristic
that led to this simple description was the approximately Gaussian
nature of the problem. Next, we attempted to present basic
classical filtering and control methodology in a self-contained,
pedagogical format. The results emphasized the inherent trade-offs
in simultaneous estimation of distinct, but dynamically coupled,
system parameters.  Because these methods are potentially critical
in any field involving optimal estimation, we consider the full
exposition of this elementary example to be a useful resource for
future analogous work.

We have also demonstrated the general principle that precision
feedback control can make estimation robust to the uncertainty of
system parameters. Despite the need to assume that the controller
produced a precise cancellation field, this approach deserves
further investigation because of its inherent ability to precisely
track broadband field signals \cite{Geremia2003b}.  It is
anticipated that these techniques will become more pervasive in
the experimental community as quantum systems are refined to
levels approaching their fundamental limits of performance.

\acknowledgments

This work was supported by the NSF (PHY-9987541, EIA-0086038), the
ONR (N00014-00-1-0479), and the Caltech MURI Center for Quantum
Networks (DAAD19-00- 1-0374). JKS acknowledges a Hertz fellowship.
The authors thank Ramon van Handel for useful discussions.
Additional information is available at
http://minty.caltech.edu/Ensemble.

\appendix

\section{Simplified representation of the plant}
\label{Appendix::SimplifiedRepresentation}

In \refsec{QuantumParameterEstimation} we outlined a general
approach to quantum parameter estimation based on the stochastic
master equation (SME), but subsequently we derived optimal
observer and controller gains from an explicit representation of
the plant dynamics (\refeqn{Dynamics}). This representation
appears classical in that the plant state is given by a scalar
variable, $z$, rather than a density operator. In this Section we
present a derivation of this simplified representation and discuss
the equivalence of our approach to the original quantum estimation
problem.

From the perspective of quantum filtering theory we will simply
show that a Gaussian approximation to the relevant SME can be
viewed as a Kalman filter, which in turn induces a simplified
representation of the dynamics of the spin state. In this
simplified representation the quantum state of the spin system is
replaced by a scalar variable, $z$, and $\langle J_z\rangle(t)$ is
viewed as the optimal estimate of the random process $z(t)$.
Equations for $dz(t)$ and its relation to the observed
photocurrent $y(t)dt$ are given in \refeqns{Simple}{Currents},
which have the convenient property of being formally
time-invariant. The technical approach in the main body of the
text is then to replace \refeqn{ReducedSME1}, which is derived
from the SME, by a state-space observer derived directly from the
simplified model of \refeqn{Simple}. By doing so we achieve
transparent correspondence with classical estimation and control
theory. We should note that the diagrams in \reffig{Equivalences}
indicate signal flows and dependencies in a way that is quite at
odds with the quantum filtering perspective. This Figure is meant
solely to motivate the simplified model (\refeqn{Simple}) for
readers who prefer a more traditional quantum optics perspective,
in which the Ito increment in the SME corresponds to optical
shot-noise (as opposed to an innovation process derived from the
photocurrent) and the SME itself plays the role of a `physical'
evolution equation mapping $h(t)$ to $y(t)dt$.

Adopting the latter perspective, let us briefly discuss (with
reference to the top diagram in \reffig{Equivalences}) the overall
structure of our estimation problem. The physical system that
exists in the laboratory (the spins and optical probe beam) acts
as a transducer, whose key role in the magnetometry scheme is to
imprint a statistical signature of the magnetic field $h(t)$ onto
the observable photocurrent, $y(t)dt$. Hence whatever theoretical
model we adopt for describing the spin and probe dynamics must
provide an accurate description of the mapping from $h(t)$ to
$y(t)dt$, as represented by the \textit{Plant} in
\reffig{Schematic}C. An open-loop estimator, designed on the basis
of this plant model, would construct a conditional probability
distribution for $h(t)$ based on passive observation of $y(t)dt$.
In a closed-loop estimation procedure we would allow the
controller to apply compensation fields to the system in order to
gain accuracy and/or robustness. In either case, the essential
role of the spin-probe (plant) model in the design process is to
provide an accurate description of the influence of an arbitrary
time-dependent field $h(t)$ on the photocurrent $y(t)dt$. Note
that the consideration of {\em arbitrary} $h(t)$ subsumes all
possible effects of real-time feedback.

\begin{figure}
\includegraphics[width=3.25in]{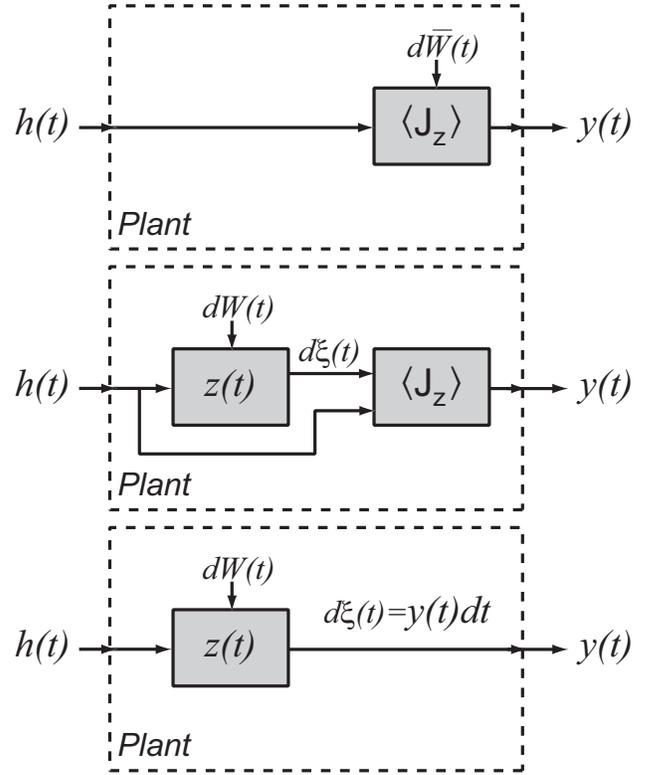}
\caption{Equivalent models for the filtering problem (see
discussion at the beginning of \refapx{SimplifiedRepresentation}).
Each version can be inserted into the \textit{Plant} block of
\reffig{Schematic} C. The filters all presume complete knowledge
of $h(t)=b(t)+u(t)$.}\label{Figure::Equivalences}
\end{figure}

Thomsen and co-workers \cite{Thomsen2002} have derived an accurate
plant model for our magnetometry problem, in the form of an SME
(\refeqn{SME}). Following a common convention in quantum optics,
let us here write this SME and the corresponding photocurrent
equation in the form
\begin{eqnarray*}
d\rho(t)&=&-i\,dt[H(h),\rho(t)]+\mathcal{D}[\sqrt{M}J_z]\rho(t)\,dt\\
&&+\sqrt{\eta}\mathcal{H}[\sqrt{M}J_z]\rho(t)d\bar{W}(t),\\
y(t)dt&=&\langle J_z\rangle(t)dt + \sqrt{\sigma_M}\,d\bar{W}(t),
\end{eqnarray*}
where $H(h)=\gamma h J_y$ and $\rho(t)$ is the state of the spin
system conditioned on the measurement record $y(t)dt$. The
quantity $d\bar{W}(t)$ is a Wiener increment that heuristically
represents shot-noise in the photodetection process
\cite{Wiseman1993}, and these are to be interpreted as Ito
stochastic differential equations. If $h(t)$ and $y(t)dt$ are
considered as input and output signals, respectively, this pair of
equations jointly implement a plant transfer function as depicted
in \reffig{Equivalences}, with $\rho(t)$ taking on the role of the
plant state.

For a large spin ensemble, however, $\rho(t)$ will have very high
dimension and it would be impractical to utilize the full SME for
design purposes. It is straightforward to derive a reduced model
by employing a moment expansion for the observable of interest.
Extracting the conditional expectation values of the first two
moments of $J_z$ from SME gives the following scalar stochastic
differential equations:
\begin{eqnarray*}
d\langle J_z \rangle(t)&=&\gamma \langle J_x \rangle(t) h(t)\,dt + \frac{\langle \Delta J_z^2\rangle(t)}{\sqrt{\sigma_M}} d\bar{W}(t)\\
d\langle \Delta J_z^2\rangle(t)&=&-\frac{\langle\Delta J_z^2\rangle^2(t)}{\sigma_M}\,dt\\
&&-i\gamma\langle[\Delta J_z^2 ,  J_y]\rangle(t)h(t)\,dt\\
&&+ \frac{\langle \Delta J_z^3 \rangle(t)}{\sqrt{\sigma_M}}
d\bar{W}(t)
\end{eqnarray*}

If the spins are initially fully polarized along $x$ and the spin
angle $\sim \langle J_z\rangle/\langle J_x\rangle$ is kept small
({\it e.g.}, by active control), then, by using the evolution
equation for the $x$ component, we can show $\langle J_x
\rangle(t) \approx J \exp[-M t /2] \approx J$ for times $t<1/M$.
Making the Gaussian approximation at small times, the third order
terms $\langle \Delta J_z^3 \rangle$ and $-i\gamma\langle[\Delta
J_z^2 , J_y]\rangle(t)h(t)$ can be neglected. The
Holstein-Primakoff transformation \cite{Holstein1940}, commonly
used in the condensed matter physics literature, makes it possible
to derive this Gaussian approximation as an expansion in $1/J$.
Both of the removed terms can be shown to be approximately
$1/J\sqrt{J}$ smaller than the retained non-linear term.
Additionally, the second removed term will be reduced if
$h(t)\approx 0$ by active control.

These approximations give
\begin{eqnarray}
d\langle J_z \rangle(t)&=&\gamma J h(t)\,dt + \frac{\langle \Delta J_z^2\rangle(t)}{\sqrt{\sigma_M}} d\bar{W}(t)\label{Equation::ReducedSME1}\\
d\langle \Delta J_z^2\rangle(t)&=&-\frac{\langle\Delta
J_z^2\rangle^2(t)}{\sigma_M}\,dt \label{Equation::ReducedSME2}
\end{eqnarray}
which constitute a Gaussian, small-time approximation to the full
SME that represents the essential dynamics for magnetometry. Note
that we can analytically solve
\begin{eqnarray*}
\langle\Delta J_z^2\rangle(t)&=&\frac{\langle\Delta
J_z^2\rangle(0)\sigma_{M}}{\sigma_M+\langle\Delta
J_z^2\rangle(0)t},
\end{eqnarray*}
where $\langle\Delta J_z^2\rangle(0)=J/2$ for an initial coherent
spin state.

At this point we may note that \refeqns{ReducedSME1}{ReducedSME2}
have the algebraic form of a Kalman filter. (This is not at all
surprising since the SME, as written in \refeqn{SME}, represents
an optimal nonlinear filter for the reduced spin state
\cite{Belavkin1999,Verstraete2001} and our subsequent
approximations have enforced both linearity and sufficiency of
second-order moments.) Viewed as such, the quantity $\langle
J_z\rangle (t)$ would represent an optimal (least square) estimate
of some underlying variable $z(t)$ based on observation of a
signal $d\xi(t)$, and $\langle\Delta J_z^2\rangle (t)$ would
represent the uncertainty (variance) of this estimate. It thus
stands to reason that we might be able to simplify our
magnetometry model even further if we could find an `underlying'
model for the evolution of $z(t)$ and $d\xi(t)$, for which our
equations derived from the SME would be the Kalman filter.

It is not difficult to do so, and indeed a very simple model
suffices:
\begin{eqnarray}
dz(t) &=& \gamma J h(t) dt\nonumber\\
d\xi(t) &=& z(t) dt + \sqrt{\sigma_M}
dW(t)\label{Equation::Simple}
\end{eqnarray}
where $dW(t)$ is an Wiener increment that is distinct from (though
related to) $d\bar{W}(t)$. In order to match initial conditions
with the equations derived from the SME, we should assume that the
expected value of $z(t=0)$ is zero and that the variance of our
prior distribution for $z(0)$ is $J/2$. Written in canonical form,
the Kalman filter for this hypothetical system is then
\begin{eqnarray*}
d\tilde{z}(t) &=& \gamma J h(t) dt +
\frac{\sigma_{zR}(t)}{\sqrt{\sigma_M}}
\frac{\left[d\xi(t)-\tilde{z}(t)dt\right]}{\sqrt{\sigma_M}},\\
d\sigma_{zR}(t) &=& -\frac{\sigma_{zR}^2(t)}{\sigma_M}\,dt.
\end{eqnarray*}
Here $\tilde{z}(t)$ is the optimal estimate of $z(t)$ and
$\sigma_{zR}(t)$ is the variance. We exactly recover the SME
model, \refeqns{ReducedSME1}{ReducedSME2}, by the identifications
\begin{eqnarray}
\tilde{z}(t) & \leftrightarrow & \langle J_z \rangle (t)\nonumber\\
\sigma_{zR}(t) & \leftrightarrow & \langle \Delta J_z^2
\rangle(t)\nonumber\\
\frac{\left[d\xi(t)-\tilde{z}(t)dt\right]}{\sqrt{\sigma_M}} &
\leftrightarrow & d\bar{W}(t).
\end{eqnarray}
It is important to note that the quantity
$\left[d\xi(t)-\tilde{z}dt \right]/\sqrt{\sigma_M}$ represents the
so-called {\em innovation process} of this Kalman filter, and it
is thus guaranteed (by least-squares optimality of the filter
\cite{Oksendal1998}) to have Gaussian white-noise statistics.
Hence we have solid grounds for identifying it with the Ito
increment appearing in the SME.

Given this insight, we see that our original magnetometry problem
can equivalently be viewed in a way that corresponds to the middle
diagram of \reffig{Equivalences}. In this version, we posit the
existence of a hidden transducer that imprints statistical
information about the magnetic field $h(t)$ onto a signal
$d\xi(t)$. A Kalman filter receives this signal, and from it
computes an estimate $\tilde{z}(t)$ as well as an innovation
process $d\bar{W}(t)$. (Note that as indicated in the diagram, the
Kalman filter will only function correctly if it `has knowledge
of' the true magnetic field $h(t)$ in the way that a physical
system would, but this is not an important point for what
follows.) According to the model equations, the Kalman filter then
emits the following signal to be received by our photo-detector:
\begin{eqnarray}
y(t)dt & = & \tilde{z}(t)dt + \sqrt{\sigma_M}\,d\bar{W}(t).
\end{eqnarray}
Note that $d\bar{W}(t)$ now appears as an internal variable to the
Kalman filter, computed from the input signal $d\xi(t)$ and the
recursive estimate $\tilde{z}(t)$, while the inherent randomness
is referred back to $dW(t)$. Although this may seem like an
unnecessarily complicated story, it should be noted that the
compound model with $z(t)$ and the Kalman filter predicts an {\em
identical} transfer function from $h(t)$ to the
experimentally-observed signal $y(t)dt$ to that of the equations
originally derived from the SME (top diagram in
\reffig{Equivalences}). Hence, for the purposes of analyzing and
designing magnetometry schemes, these are equivalent models.

Combining several definitions above we find
\begin{eqnarray}
y(t)dt & = & \tilde{z}(t)dt +
\sqrt{\sigma_M}\frac{\left[d\xi(t)-\tilde{z}(t)dt\right]}
{\sqrt{\sigma_M}}\nonumber\\
& = & d\xi(t)\label{Equation::Currents}.
\end{eqnarray}
It thus follows that in the compound model, the Kalman filter
actually implements a trivial transfer function and can in fact be
eliminated from the diagram. Doing this, we obtain the simplified
representation in the bottom diagram of \reffig{Equivalences}.
Here the perspective is to pretend that the internal dynamics of
the transducing physical system corresponds to the simplified
model (\refeqn{Simple}), since we can do so without making any
error in our description of the effect of $h(t)$ on the recorded
signal. We thus conclude that for the purposes of open- or
closed-loop estimation of $h(t)$, filters and controllers can in
fact be designed---without loss of performance---using the
simplified model (\refeqn{Simple}).

It is interesting to note that $z(t)$ can loosely be interpreted
as a `classical value' of the spin projection $J_z$. Since the
operator $J_z$ is a back-action evading observable, the continuous
measurement we consider is quantum non-demolition and its
backaction on the system state is minimal (conditioning without
disturbance). Hence if $h(t)=0$, we may think of the measurement
process as gradually `collapsing' the quantum state of the spin
system from an initial coherent state towards an eigenstate of
$J_z$; the hidden variable $z(t)$ in the simplified model
\refeqn{Simple} would then represent the eigenvalue corresponding
to the ultimate eigenstate, and $\tilde{z}(t)=\langle J_z
\rangle(t)$ in the Kalman filter would be our converging estimate
of it. (Again, this is as expected from the abstract perspective
of quantum filtering theory for open quantum systems.) Conditional
spin-squeezing in this case can then be understood as nothing more
than the reduction of our uncertainty as to the underlying value
of $z$ --- as we acquire information about $z$ through observation
of $d\xi(t)= y(t)dt$, our uncertainty $\sigma_{zR}(t)
\leftrightarrow\langle\Delta J_z^2\rangle(t)$ naturally decreases
below its initial coherent-state value of $J/2$. Still, the
quantum-mechanical nature of the spin system is not without
consequence, as it is known that continuous QND measurement
produces entanglement among the spins in the ensemble
\cite{Stockton2003}.

It seems worth commenting on the fact that \refeqn{Simple} clearly
predicts stationary statistics for the photocurrent $y(t)dt$,
whereas \refeqn{ReducedSME1} contains a time-dependent diffusion
coefficient that might color the statistics of $y(t)dt=\langle
J_z\rangle(t)dt + \sqrt{\sigma_M}\,d\bar{W}(t)$. In fact there is
no discrepancy. It is possible \cite{Wiseman1993} to derive the
second order time correlation function of the observed signal
$y(t)dt$ directly from the stochastic master \refeqn{SME},
\begin{eqnarray*}
\langle y(t) y(t+\tau) \rangle &=& (\langle J_{z}(t) J_{z}(t+\tau)
\rangle +\langle J_{z}(t+\tau) J_{z}(t) \rangle)/2
\\&&+\frac{1}{4\eta M}\delta (\tau)
\end{eqnarray*}
(This result could also be obtained from the standard input-output
theory of quantum optics.) Since the master equation results in
linear equations for the mean values $\langle J_{x}(t) \rangle$
and $\langle J_{z}(t)\rangle$ the quantum regression theorem
\cite{Walls1994} allows the correlation functions $\langle
J_{z}(t) J_{z}( t+\tau ) \rangle $ \ and $\langle J_{z}( t+\tau )
J_{z}( t) \rangle $ to be calculated explicitly. In this paper we
are most interested in the early time evolution for which we
obtain the expressions
\begin{eqnarray*}
\langle y(t) \rangle  =\langle J_{z}( t) \rangle
&=&\gamma b J t+O( t^{2})  \\
\langle y(0) y(t) \rangle -\langle y(0) \rangle \langle y(t)
\rangle &=&\frac{1}{4\eta M}\delta (t) +\langle \Delta
J_{z}^{2}\rangle (0) \\&& +O(t^{2}) .
\end{eqnarray*}
These correlation functions correspond to a white noise signal
which is a linear ramp with gradient $\gamma b J$ with a random
offset of variance $\langle \Delta J_{z}^{2}\rangle(0)$, in
perfect agreement with our simplified model \refeqn{Simple}. If
the statistics of $y(t)$ were Gaussian these first and second
order moments would be enough to characterize the signal
completely, and indeed for sufficiently large $J$ the problem does
become effectively Gaussian.

As a final comment we note that the essential step in the above
discussion is to observe that the equations for the first and
second order moments of the quantum state derived from the
stochastic master equation correspond to a Kalman filter for some
classical model of a noisy measurement. This correspondence holds
for the stochastic master equation corresponding to an arbitrary
linear quantum mechanical systems with continuous measurement of
observables that are linear combinations of the canonical
variables \cite{Doherty2003}. In the general case of measurements
that are not QND the equivalent classical model will have
noise-driven dynamical equations as well as noise on the measured
signal. The noise processes driving the dynamics and the measured
signal may also be correlated. The case of position measurement of
a harmonic oscillator shows all of these features
\cite{Doherty1999}.

\section{Riccati equation solution method}\label{Appendix::RiccatiSolution}

The matrix Riccati equation is ubiquitous in optimal control.
Here, following \cite{Reid1972}, we show how to reduce the
non-linear problem to a set of linear differential equations.
Consider the generic Riccati Equation:
\begin{equation*}
\frac{d\textbf{V}(t)}{dt}=\textbf{C}-\textbf{D}
\textbf{V}(t)-\textbf{V}(t) \textbf{A}-\textbf{V}(t) \textbf{B}
\textbf{V}(t)
\end{equation*}
We propose the decomposition:
\begin{equation*}
\textbf{V}(t)=\textbf{W}(t) \textbf{U}^{-1}(t)
\end{equation*}
with the linear dynamics
\begin{center}
\begin{tabular}[t]{c}
$\begin{bmatrix} \frac{d\textbf{W}(t)}{dt} \\
\frac{d\textbf{U}(t)}{dt}\end{bmatrix} =
\begin{bmatrix}
-\textbf{D} & \textbf{C}\\
\textbf{B} & \textbf{A}
\end{bmatrix}
\begin{bmatrix} \textbf{W}(t) \\ \textbf{U}(t)\end{bmatrix}$
\end{tabular}
\end{center}
It is straightforward to then show that this linearized solution
is equivalent to the Riccati equation
\begin{eqnarray*}
\frac{d\textbf{V}(t)}{dt}&=&\frac{d\textbf{W}(t)}{dt}\textbf{U}^{-1}+\textbf{W}(t)\frac{d\textbf{U}^{-1}(t)}{dt}\\
&=&\frac{d\textbf{W}(t)}{dt}\textbf{U}^{-1}(t)+\textbf{W}(t)(-\textbf{U}^{-1}(t)\frac{d\textbf{U}(t)}{dt}\textbf{U}^{-1}(t))\\
&=&(-\textbf{D}\textbf{W}(t)+\textbf{C}\textbf{U}(t))\textbf{U}^{-1}(t)\\
&&-\textbf{W}(t)\textbf{U}^{-1}(t)(\textbf{B}\textbf{W}(t)+\textbf{A}\textbf{U}(t))\textbf{U}^{-1}(t)\\
&=&\textbf{C}-\textbf{D}\textbf{V}(t)-\textbf{V}(t)\textbf{A}-\textbf{V}(t)\textbf{B}\textbf{V}(t)
\end{eqnarray*}
where we have used the identity
\begin{equation*}
\frac{d\textbf{U}^{-1}(t)}{dt}=-\textbf{U}^{-1}(t)\frac{d\textbf{U}(t)}{dt}\textbf{U}^{-1}(t)
\end{equation*}
Thus the proposed solution works and the problem can be solved
with a linear set of differential equations.

\section{Robust control in frequency
space}\label{Appendix::RobustFrequencySpace}

Here we apply traditional frequency-space robust control methods
\cite{Doyle1990, Doyle1997} to the classical version of our
system. This analysis is different from the treatment in the body
of the paper in several respects. First, we assume nothing about
the noise sources (bandwidth, strength, etc.). Also, this approach
is meant for steady state situations, with the resulting
estimator-controller being a constant gain transfer function. The
performance criterion we present here is only loosely related to
the more complete estimation description above. Despite these
differences, this analysis gives a very similar design procedure
for the steady state situation.

We proceed as follows with the control system shown in
\reffig{BlockDiagram}, where we label $h(t)=u(t)+ b(t)$ as the
total field. Consider the usual spin system but ignore noise
sources and assume we can measure $z(t)$ directly, so that
$z(t)=y(t)$. For small angles of rotation, the transfer function
from $h(t)$ to $y(t)$ is an integrator
\begin{eqnarray*}
\frac{dy(t)}{dt}&=&\frac{dz(t)}{dt}=\gamma J h(t)\\
s y(s)&=&\gamma J h(s)\\
y(s)&=&P(s) h(s)\\
P(s)&=&\gamma J /s
\end{eqnarray*}

\begin{figure}[b]
\includegraphics[width=3.25in]{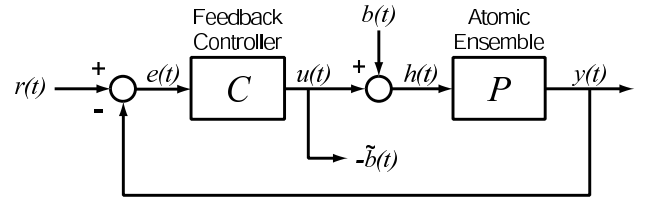} \caption{
Spin control system with plant transfer function $P(s)=\gamma
J/s$. $r(t)$ is the reference signal, which is usually zero.
$e(t)$ is the error signal. $u(t)$ is the controller output.
$b(t)$ is the external field to be tracked. $h(t)=b(t)+u(t)$ is
the total field. $\tilde{b}(t)$ is the field
estimate.}\label{Figure::BlockDiagram}
\end{figure}

Now we define the performance criterion.  First notice that the
transfer function from the field to be measured $b(t)$ to the
total field $h(t)$ is $S(s)$ where
\begin{eqnarray*}
h(s)&=&S(s)b(s)\\
S(s)&=&\frac{1}{1+P(s)C(s)}
\end{eqnarray*}
(Also notice that this represents the transfer function from the
reference to the error signal $e(s)=S(s)r(s)$.)  Because our field
estimate will be $\tilde{b}(t)=-u(t)$, we desire $h(t)$ to be
significantly suppressed.  Thus we would like $S(s)$ to be small
in magnitude (controller gain $|C(s)|$ large) in the frequency
range of interest.  However, because the gain $|C(s)|$ must
physically decrease to zero at high frequencies we must close the
feedback loop with adequate phase margin to keep the closed-loop
system stable. This is what makes the design of $C(s)$
non-trivial.

Proceeding, we now define a function $W_1(s)$ which represents the
degree of suppression we desire at the frequency $s=j\omega$.  So
our controller $C(s)$ should satisfy the following performance
criterion
\begin{eqnarray*}
\|W_1(s)S(s)\|_{\infty}&<&1\label{Equation::W1Performance}
\end{eqnarray*}
Thus the larger $W_1(s)$ becomes, the more precision we desire at
the frequency $s$.  We choose the following performance function
\begin{eqnarray*}
W_1(s)&=&\frac{W_{10}}{1+s/\omega_1}
\end{eqnarray*}
such that $\omega_1$ is the frequency below which we desire
suppression $1/W_{10}$.

Because our knowledge of $J$ is imperfect, we need to consider all
plant transfer functions in the range
\begin{eqnarray*}
P=\frac{\gamma}{s}\{J_{min}\rightarrow J_{max}\}
\end{eqnarray*}

Our goal is now to find a $C(s)$ that can satisfy the performance
condition for any plant in this family. We choose our nominal
controller as
\begin{eqnarray*}
C_0(s)&=&\frac{\omega_C}{\gamma J'}
\end{eqnarray*}
So if $J=J'$ then the system closes at $\omega_C$ (i.e.,
$|P(i\omega_C)C_0(i\omega_C)|=1$, whereas in general the system
will close at $\omega_{CR}=\omega_C\frac{J}{J'}$.  We choose this
controller because $P(s)C(s)$ should be an integrator ($\propto
1/s$) near the closing frequency for optimal phase margin and
closed loop stability.

Next we insert this solution into the performance condition.  We
make the simplifying assumption $\omega_1\ll\omega_C \frac{J}{J'}$
(we will check this later to be self-consistent). Then the optimum
of the function is obvious and the condition of
\refeqn{W1Performance} becomes
\begin{eqnarray*}
\omega_1 W_{10}<\omega_{CR}=\omega_C \frac{J}{J'}
\end{eqnarray*}
We want this condition to be satisfied for all possible spin
numbers, so we must have
\begin{eqnarray}
\omega_1 W_{10}=\min[\omega_{CR}]=\omega_C \frac{J_{min}}{J'}
\label{Equation::Jmin}
\end{eqnarray}

Experimentally, we are forced to roll-off the controller at some
high frequency that we shall call $\omega_Q$.  Electronics can
only be so fast. Of course, we never want to close above this
frequency because the phase margin would become too small, so this
determines the maximum $J$ that the controller can reliably handle
\begin{eqnarray}
\omega_Q=\max[\omega_{CR}]=\omega_C
\frac{J_{max}}{J'}\label{Equation::Jmax}
\end{eqnarray}

Combining \refeqns{Jmin}{Jmax} we find our fundamental trade-off
\begin{eqnarray}
\omega_1
W_{10}=\omega_Q\frac{J_{min}}{J_{max}}\label{Equation::W1TradeOff}
\end{eqnarray}
which is the basic result of this section.  Given experimental
constraints (such as $J_{min}$, $J_{max}$, and $\omega_Q$), it
tells us what performance to expect ($1/W_{10}$ suppression) below
a chosen frequency $\omega_1$.

From \refeqn{W1TradeOff}, we recognize that the controller gain at
the closing frequency needs to be
\begin{eqnarray*}
|C|_C=\frac{\omega_C}{\gamma J'}=\frac{\omega_1 W_{10}}{\gamma
J_{min}}=\frac{\omega_Q}{\gamma J_{max}}
\end{eqnarray*}
In the final analysis, we do not need to use $J'$ and $\omega_C$
to parametrize the controller, only the trade-off and the gain.
Also, notice that now we can express $\min[\omega_{CR}]=\omega_1
W_{10}$.

To check our previous assumption
\begin{eqnarray*}
\omega_1&\ll&\omega_C \frac{J}{J'}\\
&=&\omega_1 W_{10} \frac{J}{J_{min}}
\end{eqnarray*}
which is true if $W_{10}\gg 1$.

Finally, the system will never close below the frequency
$\min[\omega_{CR}]$ so we should increase the gain below a
frequency $\omega_H$ which we might as well set equal to
$\min[\omega_{CR}]$. This improves the performance above and
beyond the criterion above. Of course we will be forced to level
off the gain at some even lower frequency $\omega_L$ because
infinite DC gain (a real integrator) is unreasonable.  So the
final controller can be expressed as
\begin{eqnarray*}
C(s)=|C|_C
\frac{1}{1+s/\omega_Q}\frac{\omega_H(1+s/\omega_H)}{\omega_L(1+s/\omega_L)}
\end{eqnarray*}
with the frequencies obeying the order
\begin{eqnarray*}
\omega_L &<& \\
\omega_H &=& \min[\omega_{CR}]=\omega_1 W_{10} < \\
\omega_{CR} &=& \frac{J}{J_{min}}\omega_1 W_{10} < \\
\omega_Q &=& \max[\omega_{CR}]=\frac{J_{max}}{J_{min}}\omega_1
W_{10}
\end{eqnarray*}

Notice that the controller now looks like the steady state
transfer function in \reffig{Bode} derived from the steady state
of the full dynamic filter.  (The notation is the same to make
this correspondence clear).  Here $\omega_Q$ was simply stated,
whereas there it was a function of $\lambda$ that went to infinity
as $\lambda\rightarrow \infty$.  Here the high gain due to
$\omega_L$ and $\omega_H$ was added manually, whereas before it
came from the design procedure directly.

\bibliography{Paper}

\begin{thebibliography}{31}
\expandafter\ifx\csname natexlab\endcsname\relax\def\natexlab#1{#1}\fi
\expandafter\ifx\csname bibnamefont\endcsname\relax
  \def\bibnamefont#1{#1}\fi
\expandafter\ifx\csname bibfnamefont\endcsname\relax
  \def\bibfnamefont#1{#1}\fi
\expandafter\ifx\csname citenamefont\endcsname\relax
  \def\citenamefont#1{#1}\fi
\expandafter\ifx\csname url\endcsname\relax
  \def\url#1{\texttt{#1}}\fi
\expandafter\ifx\csname urlprefix\endcsname\relax\def\urlprefix{URL }\fi
\providecommand{\bibinfo}[2]{#2}
\providecommand{\eprint}[2][]{\url{#2}}

\bibitem[{\citenamefont{Armen et~al.}(2002)\citenamefont{Armen, Au, Stockton,
  Doherty, and Mabuchi}}]{Armen2002}
\bibinfo{author}{\bibfnamefont{M.~A.} \bibnamefont{Armen}},
  \bibinfo{author}{\bibfnamefont{J.~K.} \bibnamefont{Au}},
  \bibinfo{author}{\bibfnamefont{J.~K.} \bibnamefont{Stockton}},
  \bibinfo{author}{\bibfnamefont{A.~C.} \bibnamefont{Doherty}},
  \bibnamefont{and} \bibinfo{author}{\bibfnamefont{H.}~\bibnamefont{Mabuchi}},
  \bibinfo{journal}{Phys. Rev. Lett} \textbf{\bibinfo{volume}{89}},
  \bibinfo{pages}{133602} (\bibinfo{year}{2002}).

\bibitem[{\citenamefont{Geremia
  et~al.}(2003{\natexlab{a}})\citenamefont{Geremia, Stockton, and
  Mabuchi}}]{Geremia2003b}
\bibinfo{author}{\bibfnamefont{J.~M.} \bibnamefont{Geremia}},
  \bibinfo{author}{\bibfnamefont{J.~K.} \bibnamefont{Stockton}},
  \bibnamefont{and} \bibinfo{author}{\bibfnamefont{H.}~\bibnamefont{Mabuchi}},
  \bibinfo{journal}{quant-ph/0309034}  (\bibinfo{year}{2003}{\natexlab{a}}).

\bibitem[{\citenamefont{Smith et~al.}(2002)\citenamefont{Smith, Reiner, Orozco,
  Kuhr, and Wiseman}}]{Orozco2002}
\bibinfo{author}{\bibfnamefont{W.~P.} \bibnamefont{Smith}},
  \bibinfo{author}{\bibfnamefont{J.~E.} \bibnamefont{Reiner}},
  \bibinfo{author}{\bibfnamefont{L.~A.} \bibnamefont{Orozco}},
  \bibinfo{author}{\bibfnamefont{S.}~\bibnamefont{Kuhr}}, \bibnamefont{and}
  \bibinfo{author}{\bibfnamefont{H.~M.} \bibnamefont{Wiseman}},
  \bibinfo{journal}{Phys. Rev. Lett} \textbf{\bibinfo{volume}{89}},
  \bibinfo{pages}{133601} (\bibinfo{year}{2002}).

\bibitem[{\citenamefont{Morrow et~al.}(2002)\citenamefont{Morrow, Dutta, and
  Raithel}}]{Raithel2002}
\bibinfo{author}{\bibfnamefont{N.~V.} \bibnamefont{Morrow}},
  \bibinfo{author}{\bibfnamefont{S.~K.} \bibnamefont{Dutta}}, \bibnamefont{and}
  \bibinfo{author}{\bibfnamefont{G.}~\bibnamefont{Raithel}},
  \bibinfo{journal}{Phys. Rev. Lett} \textbf{\bibinfo{volume}{88}},
  \bibinfo{pages}{093003} (\bibinfo{year}{2002}).

\bibitem[{\citenamefont{Fischer et~al.}(2002)\citenamefont{Fischer, Maunz,
  Pinkse, Puppe, and Rempe}}]{Rempe2002}
\bibinfo{author}{\bibfnamefont{T.}~\bibnamefont{Fischer}},
  \bibinfo{author}{\bibfnamefont{P.}~\bibnamefont{Maunz}},
  \bibinfo{author}{\bibfnamefont{P.~W.~H.} \bibnamefont{Pinkse}},
  \bibinfo{author}{\bibfnamefont{T.}~\bibnamefont{Puppe}}, \bibnamefont{and}
  \bibinfo{author}{\bibfnamefont{G.}~\bibnamefont{Rempe}},
  \bibinfo{journal}{Phys. Rev. Lett} \textbf{\bibinfo{volume}{88}},
  \bibinfo{pages}{163002} (\bibinfo{year}{2002}).

\bibitem[{\citenamefont{Verstraete et~al.}(2001)\citenamefont{Verstraete,
  Doherty, and Mabuchi}}]{Verstraete2001}
\bibinfo{author}{\bibfnamefont{F.}~\bibnamefont{Verstraete}},
  \bibinfo{author}{\bibfnamefont{A.~C.} \bibnamefont{Doherty}},
  \bibnamefont{and} \bibinfo{author}{\bibfnamefont{H.}~\bibnamefont{Mabuchi}},
  \bibinfo{journal}{Phys. Rev. A} \textbf{\bibinfo{volume}{64}},
  \bibinfo{pages}{032111} (\bibinfo{year}{2001}).

\bibitem[{\citenamefont{Gambetta and Wiseman}(2001)}]{Gambetta2001}
\bibinfo{author}{\bibfnamefont{J.}~\bibnamefont{Gambetta}} \bibnamefont{and}
  \bibinfo{author}{\bibfnamefont{H.~M.} \bibnamefont{Wiseman}},
  \bibinfo{journal}{Phys. Rev. A} \textbf{\bibinfo{volume}{64}},
  \bibinfo{pages}{042105} (\bibinfo{year}{2001}).

\bibitem[{\citenamefont{Mabuchi}(1996)}]{Mabuchi1996}
\bibinfo{author}{\bibfnamefont{H.}~\bibnamefont{Mabuchi}},
  \bibinfo{journal}{Quantum Semiclass. Opt.} \textbf{\bibinfo{volume}{8}},
  \bibinfo{pages}{1103} (\bibinfo{year}{1996}).

\bibitem[{\citenamefont{Belavkin}(1999)}]{Belavkin1999}
\bibinfo{author}{\bibfnamefont{V.}~\bibnamefont{Belavkin}},
  \bibinfo{journal}{Rep. on Math. Phys.} \textbf{\bibinfo{volume}{43}},
  \bibinfo{pages}{405} (\bibinfo{year}{1999}).

\bibitem[{\citenamefont{Kitagawa and Ueda}(1993)}]{Kitagawa1993}
\bibinfo{author}{\bibfnamefont{M.}~\bibnamefont{Kitagawa}} \bibnamefont{and}
  \bibinfo{author}{\bibfnamefont{M.}~\bibnamefont{Ueda}},
  \bibinfo{journal}{Phys. Rev. A} \textbf{\bibinfo{volume}{47}},
  \bibinfo{pages}{5138} (\bibinfo{year}{1993}).

\bibitem[{\citenamefont{Kuzmich et~al.}(2000)\citenamefont{Kuzmich, Mandel, and
  Bigelow}}]{Kuzmich2000}
\bibinfo{author}{\bibfnamefont{A.}~\bibnamefont{Kuzmich}},
  \bibinfo{author}{\bibfnamefont{L.}~\bibnamefont{Mandel}}, \bibnamefont{and}
  \bibinfo{author}{\bibfnamefont{N.~P.} \bibnamefont{Bigelow}},
  \bibinfo{journal}{Phys. Rev. Lett.} \textbf{\bibinfo{volume}{85}},
  \bibinfo{pages}{1594} (\bibinfo{year}{2000}).

\bibitem[{\citenamefont{Stockton et~al.}(2003)\citenamefont{Stockton, Geremia,
  Doherty, and Mabuchi}}]{Stockton2003}
\bibinfo{author}{\bibfnamefont{J.~K.} \bibnamefont{Stockton}},
  \bibinfo{author}{\bibfnamefont{J.}~\bibnamefont{Geremia}},
  \bibinfo{author}{\bibfnamefont{A.~C.} \bibnamefont{Doherty}},
  \bibnamefont{and} \bibinfo{author}{\bibfnamefont{H.}~\bibnamefont{Mabuchi}},
  \bibinfo{journal}{Phys. Rev. A} \textbf{\bibinfo{volume}{67}},
  \bibinfo{pages}{022112} (\bibinfo{year}{2003}).

\bibitem[{\citenamefont{Wineland et~al.}(1994)\citenamefont{Wineland,
  Bollinger, Itano, and Heinzen}}]{Wineland1994}
\bibinfo{author}{\bibfnamefont{D.~J.} \bibnamefont{Wineland}},
  \bibinfo{author}{\bibfnamefont{J.~J.} \bibnamefont{Bollinger}},
  \bibinfo{author}{\bibfnamefont{W.~M.} \bibnamefont{Itano}}, \bibnamefont{and}
  \bibinfo{author}{\bibfnamefont{D.~J.} \bibnamefont{Heinzen}},
  \bibinfo{journal}{Phys. Rev. A} \textbf{\bibinfo{volume}{50}},
  \bibinfo{pages}{67–88} (\bibinfo{year}{1994}).

\bibitem[{\citenamefont{Smith et~al.}(2003)\citenamefont{Smith, Chaudhury, and
  Jessen}}]{Jessen2003}
\bibinfo{author}{\bibfnamefont{G.~A.} \bibnamefont{Smith}},
  \bibinfo{author}{\bibfnamefont{S.}~\bibnamefont{Chaudhury}},
  \bibnamefont{and} \bibinfo{author}{\bibfnamefont{P.~S.}
  \bibnamefont{Jessen}}, \bibinfo{journal}{J. Opt. B: Quant. Semiclass. Opt.}
  \textbf{\bibinfo{volume}{5}}, \bibinfo{pages}{323} (\bibinfo{year}{2003}).

\bibitem[{\citenamefont{Kominis et~al.}(2003)\citenamefont{Kominis, Kornack,
  Allred, and Romalis}}]{Romalis2003}
\bibinfo{author}{\bibfnamefont{I.~K.} \bibnamefont{Kominis}},
  \bibinfo{author}{\bibfnamefont{T.~W.} \bibnamefont{Kornack}},
  \bibinfo{author}{\bibfnamefont{J.~C.} \bibnamefont{Allred}},
  \bibnamefont{and} \bibinfo{author}{\bibfnamefont{M.}~\bibnamefont{Romalis}},
  \bibinfo{journal}{Nature} \textbf{\bibinfo{volume}{422}},
  \bibinfo{pages}{596} (\bibinfo{year}{2003}).

\bibitem[{\citenamefont{Budker et~al.}(2002)\citenamefont{Budker, Gawlik,
  Kimball, Rochester, Yashchuk, and Weiss}}]{Budker2002}
\bibinfo{author}{\bibfnamefont{D.}~\bibnamefont{Budker}},
  \bibinfo{author}{\bibfnamefont{W.}~\bibnamefont{Gawlik}},
  \bibinfo{author}{\bibfnamefont{D.}~\bibnamefont{Kimball}},
  \bibinfo{author}{\bibfnamefont{S.}~\bibnamefont{Rochester}},
  \bibinfo{author}{\bibfnamefont{V.}~\bibnamefont{Yashchuk}}, \bibnamefont{and}
  \bibinfo{author}{\bibfnamefont{A.}~\bibnamefont{Weiss}},
  \bibinfo{journal}{Rev. Mod. Phys.} \textbf{\bibinfo{volume}{74}},
  \bibinfo{pages}{1153} (\bibinfo{year}{2002}).

\bibitem[{\citenamefont{Geremia
  et~al.}(2003{\natexlab{b}})\citenamefont{Geremia, Stockton, Doherty, and
  Mabuchi}}]{Geremia2003}
\bibinfo{author}{\bibfnamefont{J.~M.} \bibnamefont{Geremia}},
  \bibinfo{author}{\bibfnamefont{J.~K.} \bibnamefont{Stockton}},
  \bibinfo{author}{\bibfnamefont{A.~C.} \bibnamefont{Doherty}},
  \bibnamefont{and} \bibinfo{author}{\bibfnamefont{H.}~\bibnamefont{Mabuchi}},
  \bibinfo{journal}{quant-ph/0306192}  (\bibinfo{year}{2003}{\natexlab{b}}).

\bibitem[{\citenamefont{Silberfarb and Deutsch}(2003)}]{Deutsch2003}
\bibinfo{author}{\bibfnamefont{A.}~\bibnamefont{Silberfarb}} \bibnamefont{and}
  \bibinfo{author}{\bibfnamefont{I.}~\bibnamefont{Deutsch}},
  \bibinfo{journal}{Phys. Rev. A} \textbf{\bibinfo{volume}{68}},
  \bibinfo{pages}{013817} (\bibinfo{year}{2003}).

\bibitem[{\citenamefont{Thomsen et~al.}(2002)\citenamefont{Thomsen, Mancini,
  and Wiseman}}]{Thomsen2002}
\bibinfo{author}{\bibfnamefont{L.~K.} \bibnamefont{Thomsen}},
  \bibinfo{author}{\bibfnamefont{S.}~\bibnamefont{Mancini}}, \bibnamefont{and}
  \bibinfo{author}{\bibfnamefont{H.~M.} \bibnamefont{Wiseman}},
  \bibinfo{journal}{Phys. Rev. A} \textbf{\bibinfo{volume}{65}},
  \bibinfo{pages}{061801} (\bibinfo{year}{2002}).

\bibitem[{\citenamefont{Doherty and Wiseman}(2003)}]{Doherty2003}
\bibinfo{author}{\bibfnamefont{A.~C.} \bibnamefont{Doherty}} \bibnamefont{and}
  \bibinfo{author}{\bibfnamefont{H.~M.} \bibnamefont{Wiseman}},
  \bibinfo{journal}{in preparation}  (\bibinfo{year}{2003}).

\bibitem[{\citenamefont{Jacobs}(1996)}]{Jacobs1996}
\bibinfo{author}{\bibfnamefont{O.~L.~R.} \bibnamefont{Jacobs}},
  \emph{\bibinfo{title}{Introduction to Control Theory}}
  (\bibinfo{publisher}{Oxford University Press}, \bibinfo{address}{New York},
  \bibinfo{year}{1996}), \bibinfo{edition}{2nd} ed.

\bibitem[{\citenamefont{Gardiner}(1985)}]{Gardiner2002}
\bibinfo{author}{\bibfnamefont{C.~W.} \bibnamefont{Gardiner}},
  \emph{\bibinfo{title}{Handbook of Stochastic Methods}}
  (\bibinfo{publisher}{Springer}, \bibinfo{address}{New York},
  \bibinfo{year}{1985}), \bibinfo{edition}{2nd} ed.

\bibitem[{\citenamefont{Bretthorst}(1988)}]{Bretthorst1988}
\bibinfo{author}{\bibfnamefont{G.~L.} \bibnamefont{Bretthorst}},
  \emph{\bibinfo{title}{Bayesian Spectrum Analysis and Parameter Estimation}}
  (\bibinfo{publisher}{Springer Verlag}, \bibinfo{year}{1988}).

\bibitem[{\citenamefont{Wiseman and Milburn}(1993)}]{Wiseman1993}
\bibinfo{author}{\bibfnamefont{H.~M.} \bibnamefont{Wiseman}} \bibnamefont{and}
  \bibinfo{author}{\bibfnamefont{G.~J.} \bibnamefont{Milburn}},
  \bibinfo{journal}{Phys. Rev. A} \textbf{\bibinfo{volume}{47}},
  \bibinfo{pages}{642} (\bibinfo{year}{1993}).

\bibitem[{\citenamefont{Holstein and Primakoff}(1940)}]{Holstein1940}
\bibinfo{author}{\bibfnamefont{T.}~\bibnamefont{Holstein}} \bibnamefont{and}
  \bibinfo{author}{\bibfnamefont{H.}~\bibnamefont{Primakoff}},
  \bibinfo{journal}{Phys. Rev.} \textbf{\bibinfo{volume}{58}},
  \bibinfo{pages}{1098} (\bibinfo{year}{1940}).

\bibitem[{\citenamefont{Oksendal}(1998)}]{Oksendal1998}
\bibinfo{author}{\bibfnamefont{B.}~\bibnamefont{Oksendal}},
  \emph{\bibinfo{title}{Stochastic Differential Equations}}
  (\bibinfo{publisher}{Springer Verlag}, \bibinfo{year}{1998}),
  \bibinfo{edition}{5th} ed.

\bibitem[{\citenamefont{Walls and Milburn}(1994)}]{Walls1994}
\bibinfo{author}{\bibfnamefont{D.~F.} \bibnamefont{Walls}} \bibnamefont{and}
  \bibinfo{author}{\bibfnamefont{G.~J.} \bibnamefont{Milburn}},
  \emph{\bibinfo{title}{Quantum Optics}} (\bibinfo{publisher}{Springer Verlag},
  \bibinfo{year}{1994}).

\bibitem[{\citenamefont{Doherty et~al.}(1999)\citenamefont{Doherty, Tan,
  Parkins, and Walls}}]{Doherty1999}
\bibinfo{author}{\bibfnamefont{A.~C.} \bibnamefont{Doherty}},
  \bibinfo{author}{\bibfnamefont{S.~M.} \bibnamefont{Tan}},
  \bibinfo{author}{\bibfnamefont{A.~S.} \bibnamefont{Parkins}},
  \bibnamefont{and} \bibinfo{author}{\bibfnamefont{D.~F.} \bibnamefont{Walls}},
  \bibinfo{journal}{Phys. Rev. A} \textbf{\bibinfo{volume}{60}},
  \bibinfo{pages}{2380} (\bibinfo{year}{1999}).

\bibitem[{\citenamefont{Reid}(1972)}]{Reid1972}
\bibinfo{author}{\bibfnamefont{W.~T.} \bibnamefont{Reid}},
  \emph{\bibinfo{title}{Riccati Differential Equations}}
  (\bibinfo{publisher}{Academic Press}, \bibinfo{address}{New York},
  \bibinfo{year}{1972}).

\bibitem[{\citenamefont{Doyle et~al.}(1990)\citenamefont{Doyle, Francis, and
  Tannenbaum}}]{Doyle1990}
\bibinfo{author}{\bibfnamefont{J.}~\bibnamefont{Doyle}},
  \bibinfo{author}{\bibfnamefont{B.}~\bibnamefont{Francis}}, \bibnamefont{and}
  \bibinfo{author}{\bibfnamefont{A.}~\bibnamefont{Tannenbaum}},
  \emph{\bibinfo{title}{Feedback Control Theory}}
  (\bibinfo{publisher}{Macmillan Publishing Co.}, \bibinfo{year}{1990}).

\bibitem[{\citenamefont{Zhou and Doyle}(1997)}]{Doyle1997}
\bibinfo{author}{\bibfnamefont{K.}~\bibnamefont{Zhou}} \bibnamefont{and}
  \bibinfo{author}{\bibfnamefont{J.~C.} \bibnamefont{Doyle}},
  \emph{\bibinfo{title}{Essentials of Robust Control}}
  (\bibinfo{publisher}{Prentice-Hall, Inc.}, \bibinfo{address}{New Jersey},
  \bibinfo{year}{1997}), \bibinfo{edition}{1st} ed.

\end{thebibliography}

\end{document}